\documentclass[prb,twocolumn,showpacs]{revtex4}
\usepackage{graphicx,epsf}
\newcommand{\bq}{{\bf q}}
\newcommand{\bk}{{\bf k}}
\newcommand{\bkp}{{\bf K}}
\newcommand{\bp}{{\bf p}}
\newcommand{\br}{{\bf r}}

\newcommand{\bS}{{\bf S}}
\newcommand{\bP}{{\bf P}}

\newcommand{\nphi}{{\rm N_\phi}}
\newcommand{\sz}{S_{\rm Z}}

\newcommand{\pz}{P_{\rm Z}}

\newcommand{\ez}{E_Z}
\newcommand{\ep}{E_P}

\begin{document}

\def\tende#1{\,\vtop{\ialign{##\crcr\rightarrowfill\crcr
\noalign{\kern-1pt\nointerlineskip} \hskip3.pt${\scriptstyle
#1}$\hskip3.pt\crcr}}\,}

\title{Quantum Hall ferromagnetism in graphene: an SU(4) bosonization approach}

\author{R. L. Doretto$^{1,2}$ and C. Morais Smith$^1$}

\affiliation{$^1$Institute for Theoretical Physics,
             Utrecht University,
             Postbus 80.195, 3508 TD Utrecht,
             The Netherlands \\
             $^2$Institut f\"ur Theoretische Physik,
             Universit\"at zu K\"oln,
             Z\"ulpicher Str. 77, 50937 K\"oln,
             Germany}

\begin{abstract}

We study the quantum Hall effect in graphene at filling factors $\nu =
0$ and $\nu = \pm 1$, concentrating on the quantum Hall ferromagnetic
regime, within a non-perturbative bosonization formalism. We start by
developing a bosonization scheme for electrons with two discrete
degrees of freedom (spin-1/2 and pseudospin-1/2) restricted to the
lowest Landau level. Three distinct phases are considered, namely the
so-called spin-pseudospin, spin, and pseudospin phases. The first
corresponds to a quarter-filled ($\nu =-1$) while the others to a
half-filled ($\nu = 0$) lowest Landau level. In each case, we show
that the elementary neutral excitations can be treated approximately
as a set of $n$-independent kinds of boson excitations. The boson
representation of the projected electron density, the spin,
pseudospin, and mixed spin-pseudospin density operators are derived.
We then apply the developed formalism to the effective continuous
model, which includes SU(4) symmetry breaking terms, recently proposed
by Alicea and Fisher. For each quantum Hall state, an effective
interacting boson model is derived and the dispersion relations of the
elementary excitations are analytically calculated. We propose that
the charged excitations (quantum Hall skyrmions) can be described as a
coherent state of bosons. We calculate the semiclassical limit of the
boson model derived from the SU(4) invariant part of the original
fermionic Hamiltonian and show that it agrees with the results of
Arovas and co-workers for SU(N) quantum Hall skyrmions. We briefly
discuss the influence of the SU(4) symmetry breaking terms in the
skyrmion energy.

\end{abstract}

\pacs{71.10.-w, 81.05.Uw, 73.43.-f, 73.43.Lp} \maketitle

\section{Introduction}

Graphene consists of a single atomic layer of carbon arranged in a
honeycomb lattice.\cite{ando,castroneto06} When an uniform
perpendicular magnetic field is applied, the system displays an
unconventional integer quantum Hall effect (QHE),\cite{geim,kim}
where the  Hall conductivity $\sigma_{xy} = 4(n+1/2)e^2/h$
($n$ integer) and the filling factor is defined as $\nu =
4(n+1/2)$. Such unusual behavior of $\sigma_{xy}$ is understood
within a single-particle model\cite{gusynin,castroneto06} which
shows that each Landau level in graphene is {\it approximately} four-fold
degenerate (valley, the so-called $\mathbf{K}$ and $\mathbf{K'}$
points, and electron spin).

More interesting, experiments performed at higher magnetic fields
showed new quantum Hall plateaus at $\nu =0,\,\pm 1$, and $\pm
4$,\cite{zhang06} indicating that the degeneracies of the $n=0$ and
$n=1$ Landau levels are lifted. In particular, for $\nu = \pm 4$, the
behavior of the minimum of the longitudinal resistance $R_{xx}$ in
terms of the total magnetic field suggests that here the quantum Hall
effect is due to the lifting of the spin degeneracy of the $n=1$
Landau level.\cite{zhang06} However, the origin of the plateaus at
$\nu =0$ and $\nu = \pm 1$ is not completely understood. Different
scenarios were proposed. It was suggested that the effect is due to
Coulomb interaction, which favors a quantum Hall ferromagnet ground
state.\cite{nomura06} Alicea and Fisher\cite{alicea06} proposed that
the plateaus might be related to symmetry breaking terms, such as
Zeeman and underlying lattice interactions, which give rise to a
paramagnetic phase as it occurs at $\nu = \pm 4$. An explanation based
on the so-called "magnetic catalysis" mechanism was proposed by
Gusynin {\it et al.}\cite{gusynin07} This mechanism predicts that the
long-range Coulomb interaction generates an excitonic gap, which lifts
the valley degeneracy {\it only} of the lowest Landau level. In
combination with the Zeeman splitting, the observed quantum Hall
plateaus at $\nu = 0$ and $\pm 1$ are understood. More recently,
Abanin {\it et al.}\cite{abanin07} argued that the transport response
of the quantum Hall state at $\nu =0$ is due to counter-circulating
edge states.

In this paper, we study the quantum Hall ferromagnetism in graphene
via a non-perturbative bosonization method
for the case of electrons with spin-1/2 and pseudospin-1/2
restricted to the lowest Landau level.
It constitutes a generalization of the
formalism\cite{doretto} recently proposed by one of us to study
the two-dimensional electron gas at $\nu = 1$ realized in GaAs
heterostructures.\cite{perspectives,yoshioka} Within this
formalism, the elementary neutral excitations (magnetic excitons)
and the skyrmion-antiskyrmion pair excitations of the system are
described in the same framework, namely an effective interacting
boson model. Such method is quite general and was used to
calculate spin excitations of the fractional quantum Hall systems
at $\nu = 1/3$ and $\nu = 1/5$,\cite{doretto05} as well as to
study Bose-Einstein condensation of magnetic excitons in the
bilayer quantum Hall system at total filling factor $\nu_T = 1$
(spinless case).\cite{doretto06}

Concerning the latter, the great majority of models proposed to study
this system assumes fully spin-polarized electrons. However, nuclear
magnetic resonance measurements\cite{spielman05,kumada05} indicate
that the electron spin degree of freedom might be relevant. Indeed, it
was suggested that the incompressible-compressible phase transition
observed in this system may involve a modification of the spin
polarization.\cite{spielman05} Therefore, theoretical tools which
allow us to properly treat the electron-electron interaction and
simultaneously take into account the electron spin and layer
(pseudospin) degrees of freedom are needed. The formalism developed
here might be also useful to study the bilayer quantum Hall system at
$\nu_T=1$ in GaAs heterostructures (spinfull case).

Our paper is organized as follows. In Sec. II, we define the
creation and annihilation boson operators and derive the boson
representation of the (projected) electron density, spin,
pseudospin, and mixed spin-pseudospin density operators. Three
distinct cases are considered, the so-called spin-pseudospin
phase, which occurs when the lowest Landau level is
quarter-filled, the spin and pseudospin phases, which are related
to a half-filled lowest Landau level. In Sec. III, we apply the
generalized bosonization formalism to study the QHE in graphene at
$\nu =0$ and $\nu = \pm 1$, focusing on the
quantum Hall ferromagnetic regime. Our starting point is the
effective continuous model recently proposed by Alicea and
Fisher.\cite{alicea06} For each quantum Hall state, an effective
interacting boson model is derived and the dispersion relations of
the elementary neutral excitations are analytically calculated. We
comment on some possible effects of the boson-boson interaction and
show how the quantum Hall skyrmion might be described within this
scheme. A summary of the main results is presented in Sec. IV.

\section{The bosonization method}
\label{method}

In order to develop a bosonization scheme for electrons with two
discrete degrees of freedom (spin-1/2 and pseudospin-1/2) and restricted
to the lowest Landau level subspace, we follow the lines of
Ref.\ \onlinecite{doretto} and start by studying the
corresponding noninteracting model.

Let us consider ${\rm N}$ noninteracting electrons moving in the
$xy$ plane under a perpendicular magnetic field ${\bf B} =
B\hat{z}$. In addition to the electronic spin ($\sigma$, $\lambda = \uparrow$,
$\downarrow$), let us also
include a discrete pseudospin index $\alpha$, $\beta = \pm$. Restricting
the Hilbert space to the lowest Landau level, the kinetic energy is
quenched and therefore the Hamiltonian of
the system is
\begin{eqnarray}
\mathcal{H} &=& \mathcal{H_Z} + \mathcal{H_{PZ}} \label{freehamiltonian} \\
            &=& -\frac{1}{2}\sum_{\sigma,\alpha}\int d^2r\;\left(\sigma\ez + \alpha\ep\right)
                \Psi^\dagger_{\alpha\,\sigma}(\br)\Psi_{\alpha\,\sigma}(\br).
\nonumber
\end{eqnarray}
In addition to the Zeeman term $\mathcal{H_Z}$, we also include an
extra term ($\mathcal{H_{PZ}}$) which breaks the pseudospin
degeneracy. As we will see below, a finite $\ep$ helps us to
define a set of different reference states.
$\ez = g\mu_BB$ is the Zeeman energy, where $g$ is the effective electron
$g$-factor and $\mu_B$ is the Bohr magneton (see Appendix
\ref{appendix1}). $\Psi^\dagger_{\alpha\,\sigma}(\br)$ is a
fermion field operator that can be expanded in the (Schr\"odinger)
lowest Landau level basis $|n=0\, m\, \rangle$ (symmetric
gauge)\cite{doretto} as
\begin{eqnarray}
\Psi^\dagger_{\alpha\,\sigma}(\br) &=& \sum_m \langle n=0\, m\,
|\br\rangle c^\dagger_{m\, \alpha\, \sigma}, \nonumber \\
&& \nonumber \\
\Psi_{\alpha\,\sigma}(\br) &=& \sum_m \langle \br | n=0\, m\, \rangle
c_{m\, \alpha\, \sigma}. \label{fermionfields}
\end{eqnarray}
The operator $c^\dagger_{m\, \alpha\, \sigma}$ ($c_{m\, \alpha\,
\sigma}$) creates (destroys) an electron in the lowest Landau
level, with guiding center $m$, pseudospin $\alpha$, and spin
$\sigma$. Substituting Eq. (\ref{fermionfields}) into Eq.
(\ref{freehamiltonian}), one sees that the Hamiltonian
$\mathcal{H}$ is diagonal in the lowest Landau level basis, i.e.,
\begin{equation}
\mathcal{H} =  -\frac{1}{2}\sum_{m=0}^{\nphi-1}\sum_{\alpha,\sigma}
    \left(\sigma\ez + \alpha\ep \right)
    c^\dagger_{m\,\alpha\,\sigma}c_{m\,\alpha\,\sigma}.
\label{freehamiltonian2}
\end{equation}
The above Hamiltonian has four highly degenerate energy levels, whose
energies are
$-(\ez+\ep)/2$, $-(\ez - \ep)/2$, $(\ez -\ep)/2$ and $(\ez +\ep)$,
and the degeneracy of each level is $\nphi =
1/2\pi l^2$. Here, $l=\sqrt{\hbar c/eB}$ is the magnetic length and we
assume that the total area of the system is one.
In the following, we will concentrate on three distinct
configurations of the system: total number of electrons ${\rm N} =
\nphi$ and $\ez > \ep$, which we call spin-pseudospin phase; ${\rm N}
= 2\nphi$ and $\ez > \ep$ (spin phase);
and  ${\rm N} = 2\nphi$ and $\ez < \ep$ (pseudospin phase).

As discussed in Ref. \onlinecite{doretto}, the creation and
annihilation boson operators are defined by considering the
neutral (particle-hole) excitations above a well-defined reference
state. As each one of the above phases has a different reference
state (noninteracting ground state), the three cases will be
analyzed separately. However, before doing that, we should firstly
discuss the representation and the algebra of the electron
density, spin, pseudospin and mixed spin-pseudospin density
operators projected into the lowest Landau level.

\subsection{Density operators and the lowest Landau level algebra}

We start by defining the following projected density operator
\begin{equation}
\rho_{\alpha\sigma,\beta\lambda}(\br) =
\Psi^\dagger_{\alpha\sigma}(\br)\Psi_{\beta\lambda}(\br),
\end{equation}
where the fermion field operators are given by
Eq. (\ref{fermionfields}), and whose Fourier transform is
\begin{eqnarray}
\rho_{\alpha\sigma,\beta\lambda}(\bq)
   &=& \int
        d^2r\;e^{-i\mathbf{q}\cdot\mathbf{r}}\Psi^\dagger_{\alpha\sigma}(\br)\Psi_{\beta\lambda}(\br)
        \nonumber \\ && \nonumber \\
   &=& \sum_{m,m'}\int d^2r\;e^{-i\mathbf{q}\cdot\mathbf{r}}\langle
      m|\mathbf{r}\rangle
     \langle \mathbf{r}|m'\rangle c^\dagger_{m\, \alpha\, \sigma}c_{m'\, \beta\, \lambda}
  \nonumber \\
  && \nonumber \\
   &=& e^{-(lq)^2/2}\sum_{m,m'}
      G_{m,m'}(l\bq)c^\dagger_{m\, \alpha\, \sigma}c_{m'\, \beta\, \lambda},
\label{fourier}
\end{eqnarray}
with $ q = |\bq|$. The function $G_{m,m'}(x)$ is defined as
\begin{eqnarray}
G_{m,m'}(l\bq) &=&
\theta(m'-m)\sqrt{\frac{m!}{m'!}}\left(\frac{-il(q_x -
iq_y)}{\sqrt{2}}\right)^{m'-m}
\nonumber \\
               && \times L^{m'-m}_{m}\left((lq)^2/2\right)
\nonumber \\
    &+& \theta(m-m')\sqrt{\frac{m'!}{m!}}\left(\frac{-il(q_x+iq_y)}{\sqrt{2}}\right)^{m-m'}
\nonumber \\
        && \times L^{m-m'}_{m'}\left((lq)^2/2\right),
\label{gdef}
\end{eqnarray}
where $L^{m-m'}_{m'}(x)$ is the generalized Laguerre polynomial.\cite{arfken}
Due to the fact that the operators $\rho_{\alpha\sigma,\beta\lambda}(\bq)$ are
projected into the lowest Landau level, their commutation relations are
modified, i.e,
\begin{eqnarray}
\nonumber
 \left[\rho_{\alpha\sigma,\beta\lambda}(\bq)\right. &,&\left.\rho_{\alpha'\sigma',\beta'\lambda'}(\bq')\right] =
  e^{\bq\cdot\bq'l^2/2}\\
\nonumber && \\
\nonumber
  &\times&\left[\delta_{\beta,\alpha'}\delta_{\lambda,\sigma'}
   e^{i\bq\wedge\bq'/2}\rho_{\alpha\sigma,\beta'\lambda'}(\bq+\bq') \right.\\
\nonumber && \\
  &-&\left.
     \delta_{\alpha,\beta'}\delta_{\sigma,\lambda'}e^{-i\bq\wedge\bq'/2}
     \rho_{\alpha'\sigma',\beta\lambda}(\bq+\bq')\right],
\label{commutator1}
\end{eqnarray}
where $\bq\wedge\bk \equiv l^2(\bq\times\bk)\cdot\hat{z}$.

It is convenient to introduce an isospin index $I$ such that
\[
 I = (\alpha,\,\sigma) = (+,\uparrow),\, (+,\downarrow),\, (-,\uparrow),\, (-,\downarrow) =
             1,2,3,4.
\]
In this new representation, the commutator
(\ref{commutator1}) simply reads
\begin{eqnarray}
\nonumber
[\rho_{IJ}(\bq),\rho_{\bar{I}\bar{J}}(\bq')] &=&
e^{\bq\cdot\bq'l^2/2}
\left[\delta_{J,\bar{I}}e^{i\bq\wedge\bq'/2}\rho_{I\bar{J}}(\bq+\bq')\right.\\
\nonumber \\
&&-\left.
 \delta_{I,\bar{J}}e^{-i\bq\wedge\bq'/2}\rho_{\bar{I}J}(\bq+\bq')\right].
\label{commutator}
\end{eqnarray}
It is also useful to define a four-component spinor $\hat{\Psi}^\dagger(\br)$ as
\begin{eqnarray}
\hat{\Psi}^\dagger(\br) &=&
\left(\Psi^\dagger_{+\uparrow}(\br)\;\;\Psi^\dagger_{+\downarrow}(\br)\;\;
      \Psi^\dagger_{-\uparrow}(\br)\;\;\Psi^\dagger_{-\downarrow}(\br)
\right),
\end{eqnarray}
which, in the isospin language, assumes the form
\begin{eqnarray}
\hat{\Psi}^\dagger(\br) &=&
\left(\Psi^\dagger_{1}(\br)\;\;\Psi^\dagger_{2}(\br)\;\;\Psi^\dagger_{3}(\br)\;\;\Psi^\dagger_{4}(\br)
\right). \label{spinor}
\end{eqnarray}

The (projected) electron density operator can now be written in
terms of the spinor (\ref{spinor}) as
\begin{equation}
\rho(\br) = \hat{\Psi}^\dagger(\br)\hat{\Psi}(\br) =
            \sum_{I=1}^4\Psi^\dagger_I(\br)\Psi_I(\br)
\end{equation}
and therefore its Fourier transform may be expressed in terms of
the density operators $\rho_{IJ}(\bq)$ as
\begin{equation}
 \rho(\bq) = \left[\rho_{11}(\bq) + \rho_{22}(\bq)
                   + \rho_{33}(\bq) + \rho_{44}(\bq)\right].
\label{totaldensity}
\end{equation}
The same can be done for the spin-$\sigma$ and pseudospin-$\alpha$
electron density operators
\begin{eqnarray}
\nonumber
  \rho_\uparrow(\bq) &=& \rho_{11}(\bq) + \rho_{33}(\bq), \;\;\;\;
  \rho_\downarrow(\bq) = \rho_{22}(\bq) + \rho_{44}(\bq),\\
\label{densityop} && \\
\nonumber
  \rho_+(\bq) &=& \rho_{11}(\bq) + \rho_{22}(\bq), \;\;\;\;
  \rho_-(\bq) = \rho_{33}(\bq) + \rho_{44}(\bq).
\end{eqnarray}

The definition (\ref{spinor}) implies that the structure
of the spin-pseudospin space is $SU(2)_{PS}\otimes SU(2)_{SPIN}$ and
therefore, the spin, pseudospin, and mixed spin-pseudospin density operators ($\hbar = 1$)
are respectively defined as
\begin{equation}
  \bS(\br) = \frac{1}{2}\hat{\Psi}^\dagger(\br)\left({\mathbf 1}_{2\times
  2}\otimes\hat{\sigma}\right)\hat{\Psi}(\br),
\label{spinop}
\end{equation}
\begin{equation}
\bP(\br)  = \frac{1}{2}\hat{\Psi}^\dagger(\br)\left(\hat{\sigma}
            \otimes {\mathbf 1}_{2\times 2}\right)\hat{\Psi}(\br),
\label{ppinop}
\end{equation}
\begin{equation}
\bP\bS(\br)  = \frac{1}{2}\hat{\Psi}^\dagger(\br)\left(\hat{\sigma}
            \otimes \hat{\sigma}\right)\hat{\Psi}(\br),
\label{psinop}
\end{equation}
Here, ${\mathbf 1}_{2\times 2}$ is the two-dimensional unit
matrix and $\hat{\sigma} = (\sigma_x\;\;\sigma_y\;\;\sigma_z)$ is a
vector whose components are the Pauli matrices. Expanding the Fourier
transform of the components of $\bS(\br)$ and
$\bP(\br)$ in terms of the density operators $\rho_{IJ}(\bq)$, we
have
\[
\sz(\bq) = \frac{1}{2}\left[\rho_{11}(\bq) - \rho_{22}(\bq)
         + \rho_{33}(\bq) - \rho_{44}(\bq)\right],
\]
\begin{equation}
 S^+(\bq) = [S_{\rm X}(\bq) + i S_{\rm Y}(\bq)] =
          \rho_{12}(\bq)+\rho_{34}(\bq),
\label{spinop2}
\end{equation}
\[
 S^-(\bq) = [S_{\rm X}(\bq) - i S_{\rm Y}(\bq)] =
             \rho_{21}(\bq)+\rho_{43}(\bq),
\]
and
\[
\pz(\bq) = \frac{1}{2}\left[\rho_{11}(\bq) + \rho_{22}(\bq) -
           \rho_{33}(\bq) - \rho_{44}(\bq)\right],
\]
\begin{equation}
 P^+(\bq) 
                   =\rho_{13}(\bq) + \rho_{24}(\bq),
\label{ppinop2}
\end{equation}
\[
P^-(\bq) 
         = \rho_{31}(\bq) + \rho_{42}(\bq).\nonumber
\]
Similar considerations hold for the mixed operators $\bP\bS(\br)$,
in particular, we have
\begin{equation}
\pz\sz(\bq) = \frac{1}{2}\left[\rho_{11}(\bq) - \rho_{22}(\bq) -
           \rho_{33}(\bq) + \rho_{44}(\bq)\right].
\label{psinop2}
\end{equation}
This component of $\bP\bS(\bq)$ will be important in the next sections.
We should mention that the representation
(\ref{spinop})-(\ref{psinop}) does not correspond to the standard
representation of the special unitary group SU(4) [see Ref.\
\onlinecite{greiner} for details], but it
follows the ideas presented in the Appendix A of Ref.\
\onlinecite{ezawa04}.

Finally, with the aid of the commutator (\ref{commutator}),
a long but straightforward calculation shows
that the density operators (\ref{totaldensity}), (\ref{spinop})
and (\ref{ppinop}) obey the lowest Landau level algebra (the same
results have been derived in a more general way\cite{mark-notes})
\begin{eqnarray}
\left[\rho(\bq),\rho(\bk)\right] &=&
      2i\sin\left(\bq\wedge\bk/2\right)e^{\bq\cdot\bk/2}\rho(\bq+\bk),
\nonumber \\
\nonumber && \\
\left[I^\mu_a(\bq),\rho(\bk)\right] &=&
      2i\sin\left(\bq\wedge\bk/2\right)e^{\bq\cdot\bk/2}I^\mu_a(\bq+\bk),
\nonumber \\
\nonumber && \\
 \left[I^\mu_a(\bq),I^\mu_b(\bk)\right] &=&
      (i/2)\delta_{a,b}\sin\left(\bq\wedge\bk/2\right)e^{\bq\cdot\bk/2}\rho(\bq+\bk)
\nonumber \\
\nonumber \\
      &+&
      i\epsilon^{abc}\cos\left(\bq\wedge\bk/2\right)e^{\bq\cdot\bk/2}I^\mu_c(\bq+\bk),
\nonumber \\
\nonumber && \\
\left[P_a(\bq),S_b(\bk)\right] &=&
      i\sin\left(\bq\wedge\bk/2\right)e^{\bq\cdot\bk/2}P_aS_b(\bq+\bk).
\label{algebra}
\end{eqnarray}
Here, $a,b,c = {\rm X,Y,Z}$ and $\epsilon^{abc}$ is the Levi-Civita
tensor.\cite{arfken} $\mu = S,P$  and therefore $I^S_a(\bq)$ and
$I^P_a(\bq)$
stand respectively for $S_a(\bq)$ and $P_a(\bq)$.
Due to the fact that the density operators (\ref{totaldensity}),
(\ref{spinop}), and (\ref{ppinop}) are projected into the lowest Landau
level, the algebra (\ref{algebra}) is different from
the usual one of the generators of the SU(4) group.
\cite{greiner,ezawa04}

\subsection{Spin-pseudospin polarized state}
\label{spfm-phase}

\begin{figure}[t]
\centerline{\includegraphics[height=4.0cm]{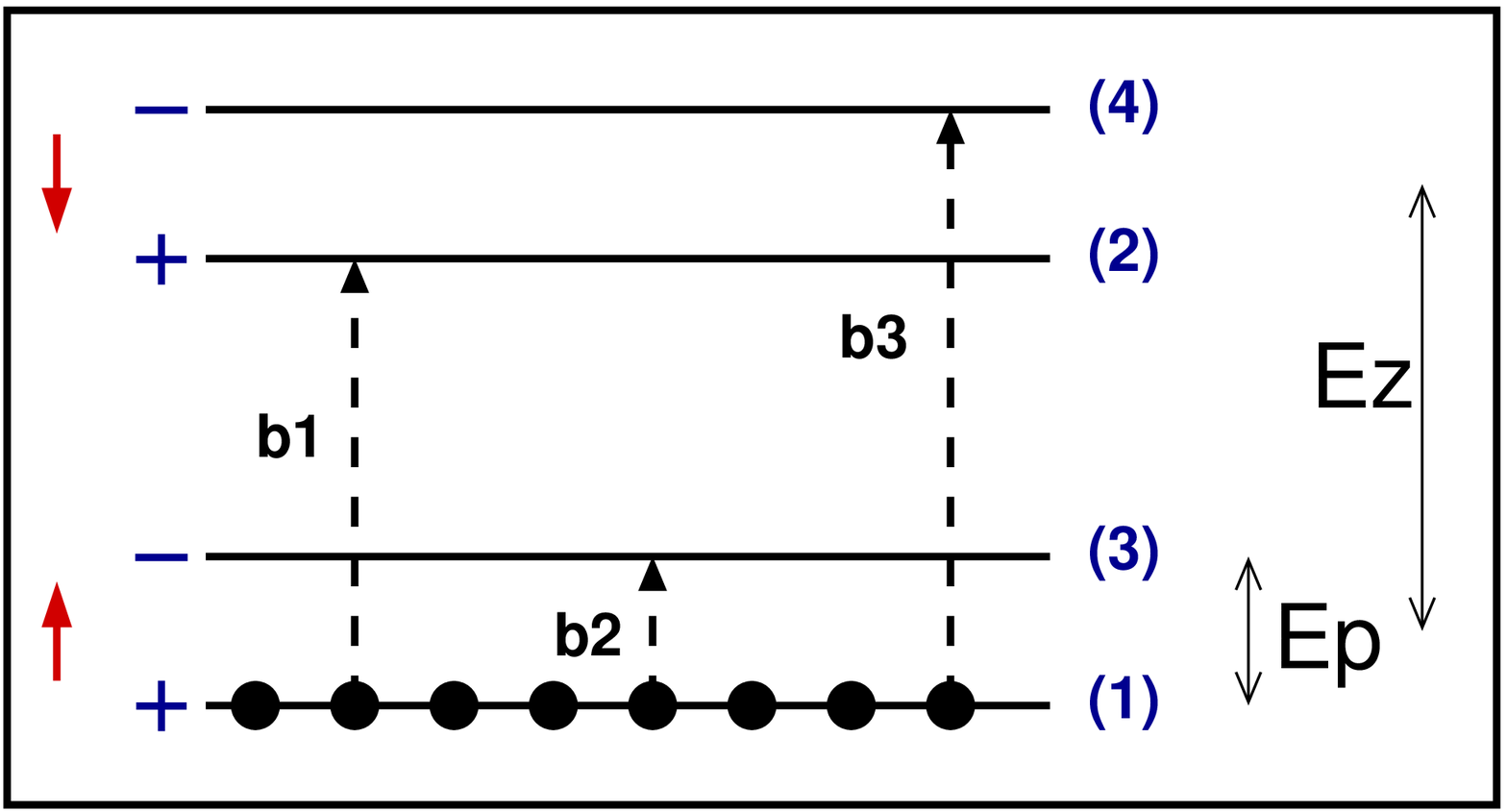}}
\caption{\label{fig01}{(color online) Schematic representation of the four highly
    degenerate lowest Landau levels when $\ez > \ep$. The
    state $|{\rm SPFM} \rangle$ is obtained by completely filling the
    energy level (1). b1, b2 and b3 are
    the elementary neutral excitations which are related to the
    density operators  $\rho_{21}(\bq)$, $\rho_{31}(\bq)$,
    and $\rho_{41}(\bq)$, respectively.}}
\end{figure}

Let us now study the noninteracting system described by the Hamiltonian
(\ref{freehamiltonian2}), assuming that the total number of
electrons ${\rm N} = \nphi$ and $\ez > \ep$. The four (highly
degenerate) energy levels are schematically displayed in Fig.\
\ref{fig01}. In this case, the noninteracting ground state of the
system is a spin-polarized pseudospin-polarized state,
\begin{equation}
|{\rm SPFM} \rangle =
 \prod_{m=0}^{\nphi-1}c^\dagger_{m\,+\,\uparrow}|0\rangle,
\label{spfm}
\end{equation}
where $|0\rangle$ is the fermion vacuum. Notice that
the neutral (particle-hole) excitations are created by
applying
the density operators $\rho_{21}(\bq)$, $\rho_{31}(\bq)$,
and $\rho_{41}(\bq)$ on the state $|{\rm SPFM} \rangle$.

From Eq. (\ref{commutator}), it follows that the commutator
between each one of the above density operators and its respective
Hermitian conjugate is
\begin{eqnarray}
\nonumber
[\rho_{1I}(\bq),\rho_{I1}(\bq')] &=&
e^{\bq\cdot\bq'l^2/2}
\left[e^{i\bq\wedge\bq'/2}\rho_{11}(\bq+\bq')\right.\\
\nonumber \\
&&-\left.
 e^{-i\bq\wedge\bq'/2}\rho_{II}(\bq+\bq')\right],
\label{commutator2}
\end{eqnarray}
with $I=2, 3, 4$. By expanding the density operators
$\rho_{II}(\bq)$ around the (reference) state (\ref{spfm}),
\begin{eqnarray}
\rho_{II}(\bq) &=& \langle {\rm SPFM}|\rho_{II}(\bq)|{\rm SPFM}\rangle +
\delta\rho_{II}(\bq)
\nonumber \\
\nonumber && \\
&=& \nphi\delta_{I,1}\delta_{\bq,0} + \delta\rho_{II}(\bq),
\label{assumption1}
\end{eqnarray}
and neglecting the fluctuations with respect to the average value,
the commutator (\ref{commutator2}) assumes the form
\begin{equation}
\nonumber [\rho_{1I}(\bq),\rho_{I1}(\bq')] \approx
\delta_{\bq,-\bq'}\nphi e^{(ql)^2/2}. \label{commutator3}
\end{equation}
One can see that, although the relations (\ref{commutator2}) do not
correspond to the usual canonical commutation relation between the
annihilation and  creation boson operators, their expectation values
in the ground state $|{\rm SPFM} \rangle$ do. In other words, as long
as the number of particle-hole excitations in the system is small,
i.e., $\langle\rho_{11}(\bq)\rangle\;\gg\;\delta\rho_{11}(\bq)$, the
density operators $\rho_{21}(\bq)$, $\rho_{31}(\bq)$, and
$\rho_{41}(\bq)$ may be approximately considered as boson operators.
Moreover, by noticing that
\[
\rho_{32}(\bq) = \rho_{42}(\bq) = \rho_{43}(\bq) \approx 0,
\]
which is related to the fact that the average values of the above
density operators with respect to the state defined by Eq.
(\ref{spfm}) vanish, it turns out that the three kinds of boson
operators are independent.

Based on the above analysis, we define the following set of
creation and annihilation boson operators
\begin{eqnarray}
b^\dagger_1(\bq)&\equiv&\alpha_q\rho_{21}(\bq),
         \;\;\;\;\;\;
b_1(\bq) \equiv \alpha_q\rho_{12}(-\bq),
\nonumber \\
b^\dagger_2(\bq)&\equiv&\alpha_q\rho_{31}(\bq),
         \;\;\;\;\;\;
b_2(\bq) \equiv \alpha_q\rho_{13}(-\bq),
\label{bosons1} \\
b^\dagger_3(\bq)&\equiv&\alpha_q\rho_{41}(\bq),
         \;\;\;\;\;\;
b_3(\bq) \equiv \alpha_q\rho_{14}(-\bq),
\nonumber
\end{eqnarray}
with $\alpha_q = e^{(lq)^2/4}/\sqrt{\nphi}$. From now on, we will
assume that the above operators obey the usual canonical algebra
\[
  [b^\dagger_i(\bq), b^\dagger_j(\bk)] =
  [b_i(\bq), b_j(\bk)] = 0,
\]
\begin{equation}
  [b_i(\bq), b^\dagger_j(\bk)] = \delta_{i,j}\delta_{\bq,\bk}.
\label{boson-algebra}
\end{equation}
Finally, we should mention that the reference state $|{\rm SPFM}
\rangle$ is indeed the boson vacuum as one can easily show that
$b_i(\bq)|{\rm SPFM} \rangle = 0$.

Once the boson operators are defined, the boson representation of
any operator $\mathcal{O}$ is determined by examining the
commutators $[\mathcal{O}, b^\dagger_i(\bk)]$ ($i=1,2,3$) and the
action of $\mathcal{O}$ in the reference state $|{\rm SPFM}
\rangle$. For instance, let us consider the density operator
$\rho_{11}(\bq)$. From Eqs.(\ref{commutator}) and (\ref{bosons1}),
we have
\[
  [\rho_{11}(\bq),b^\dagger_i(\bk)]
          = -e^{-(lq)^2/4}e^{-i\bq\wedge\bk/2}b^\dagger_i(\bq+\bk),
\]
with $i=1,2,3$. Moreover,
\[
\rho_{11}(\bq)|{\rm SPFM}\rangle = \nphi\delta_{\bq,0}|{\rm
SPFM}\rangle.
\]
Using the fact that the three kinds of boson operators
(\ref{bosons1}) are independent, the above relations are satisfied
if the density operator $\rho_{11}(\bq)$ is expanded in terms of
the bosons $b_i(\bq)$ as
\begin{equation}
\rho_{11}(\bq) = \nphi\delta_{\bq,0} - e^{-(lq)^2/4}
                 \sum_{\bk,i} e^{-i\bq\wedge\bk/2}
                 b^\dagger_i(\bq+\bk)b_i(\bk).
\label{boson-density1}
\end{equation}
Similarly, it is possible to show that
\begin{eqnarray}
\nonumber \rho_{22}(\bq) &=&
                e^{-(lq)^2/4}\sum_\bk e^{i\bq\wedge\bk/2}
                b^\dagger_1(\bq+\bk)b_1(\bk), \\
\label{boson-density2} \rho_{33}(\bq) &=&
                e^{-(lq)^2/4}\sum_\bk e^{i\bq\wedge\bk/2}
                b^\dagger_2(\bq+\bk)b_2(\bk), \\
\nonumber \rho_{44}(\bq) &=&
                e^{-(lq)^2/4}\sum_\bk e^{i\bq\wedge\bk/2}
                b^\dagger_3(\bq+\bk)b_3(\bk),
\end{eqnarray}
i.e., the expansions of all density operators $\rho_{II}(\bq)$ in
terms of bosons are {\it quadratic}.

With the aid of the relations (\ref{boson-density1}) and
(\ref{boson-density2}), one can easily write down the boson
representation of the electron density [Eq. (\ref{totaldensity})],
the $z$-components of the spin [Eq. (\ref{spinop2})] and
pseudospin [Eq. (\ref{ppinop2})] density operators, and the mixed
spin-pseudospin density operator $P_ZS_Z(\bq)$ [Eq. (\ref{psinop2})],
namely
\begin{eqnarray}
\nonumber
\rho(\bq) &=& \nphi\delta_{\bq,0} + 2ie^{-(lq)^2/4} \\
\nonumber && \\
          &&  \times\sum_{i,\bk}
\label{densityop-boson}  \sin\left(\bq\wedge\bk/2\right)
              b^\dagger_i(\bq+\bk)b_i(\bk), \\
\nonumber && \\
\nonumber I^\mu_Z(\bq) &=& \frac{1}{2}\nphi\delta_{\bq,0} + \sum_{i,\bk}
              f^\mu_i(\bq,\bk)
              b^\dagger_i(\bq+\bk)b_i(\bk),\\
\label{ipinop-boson}
\end{eqnarray}
with $I^\mu_Z(\bq) = \sz(\bq), \pz(\bq)$, and $\sz\pz(\bq)$, and
the form factors $f^\mu_i(x)$ are given by
\begin{eqnarray}
\nonumber f^S_1(\bq,\bk) &=&
          f^S_3(\bq,\bk) =
          -e^{-(lq)^2/4}\cos\left(\bq\wedge\bk\right/2), \\
\nonumber f^S_2(\bq,\bk) &=&
          ie^{-(lq)^2/4}\sin\left(\bq\wedge\bk\right/2),
\end{eqnarray}
\begin{eqnarray}
\label{formfactors-spfm}
          f^P_1(\bq,\bk) &=&
          ie^{-(lq)^2/4}\sin\left(\bq\wedge\bk\right/2), \\
\nonumber
          f^P_2(\bq,\bk) &=&
          f^P_3(\bq,\bk) =
          -e^{-(lq)^2/4}\cos\left(\bq\wedge\bk\right/2),
\end{eqnarray}
and
\begin{eqnarray}
\nonumber f^{PS}_1(\bq,\bk) &=&
          f^{PS}_2(\bq,\bk) =
          -e^{-(lq)^2/4}\cos\left(\bq\wedge\bk\right/2), \\
\nonumber f^{PS}_3(\bq,\bk) &=&
          ie^{-(lq)^2/4}\sin\left(\bq\wedge\bk\right/2).
\end{eqnarray}

In addition to the set of density operators analyzed above, the boson
representation of the density operators $\rho_{21}(\bq)$ and
$\rho_{34}(\bq)$ and the respective Hermitian conjugates
$\rho_{12}(-\bq)$ and $\rho_{43}(-\bq)$ will be useful in the next
section, where the bosonization scheme will be applied to study the
QHE in graphene. Sometimes, the expressions are not so simple as the
one presented above [see Appendix \ref{appendix2}]. Another important
point is that sometimes the Hermiticity requirement is not
full-filled. For instance, in the spin-pseudospin phase, the expansion
of $\rho_{21}(\bq)$ in terms of bosons does not correspond to the one
of $\rho_{12}(-\bq)$. As it was already discussed in Ref.
\onlinecite{doretto}, it does not constitute a major problem because
the boson expressions (\ref{densityop-boson}), (\ref{ipinop-boson}),
(\ref{ops1-spfm}) and (\ref{ops2-spfm}), derived within the procedure
outlined above satisfy the lowest Landau level algebra
(\ref{algebra}).

The asymmetric boson representation found for some operators
might be related to the fact that the bosonization method
explicitly breaks some symmetries. For instance, in the
spin-pseudospin phase, the spin "directions" up and down are no
longer equivalent because the bosons $b_i(\bq)$ are defined with
respect to the reference state $|{\rm SPFM}\rangle$. As a consequence,
the  bosonic expressions of the spin density operators $S^-(\bq) =
\rho_{21}(\bq) + \rho_{43}(\bq)$ and
$S^+(\bq) = \rho_{12}(\bq) + \rho_{34}(\bq)$ [see Eq.
(\ref{spinop2})] are asymmetric. We will see later in
Sec. \ref{pfm-phase} that the bosonic representation of the spin
density operators satisfies the condition
$S^+(\bq) = [S^-(-\bq)]^\dagger$ because for the
pseudospin phase {\it only} the pseudospin
symmetry is explicitly broken.

\subsection{Spin phase}
\label{sfm-phase}

\begin{figure}[t]
\centerline{\includegraphics[height=4.0cm]{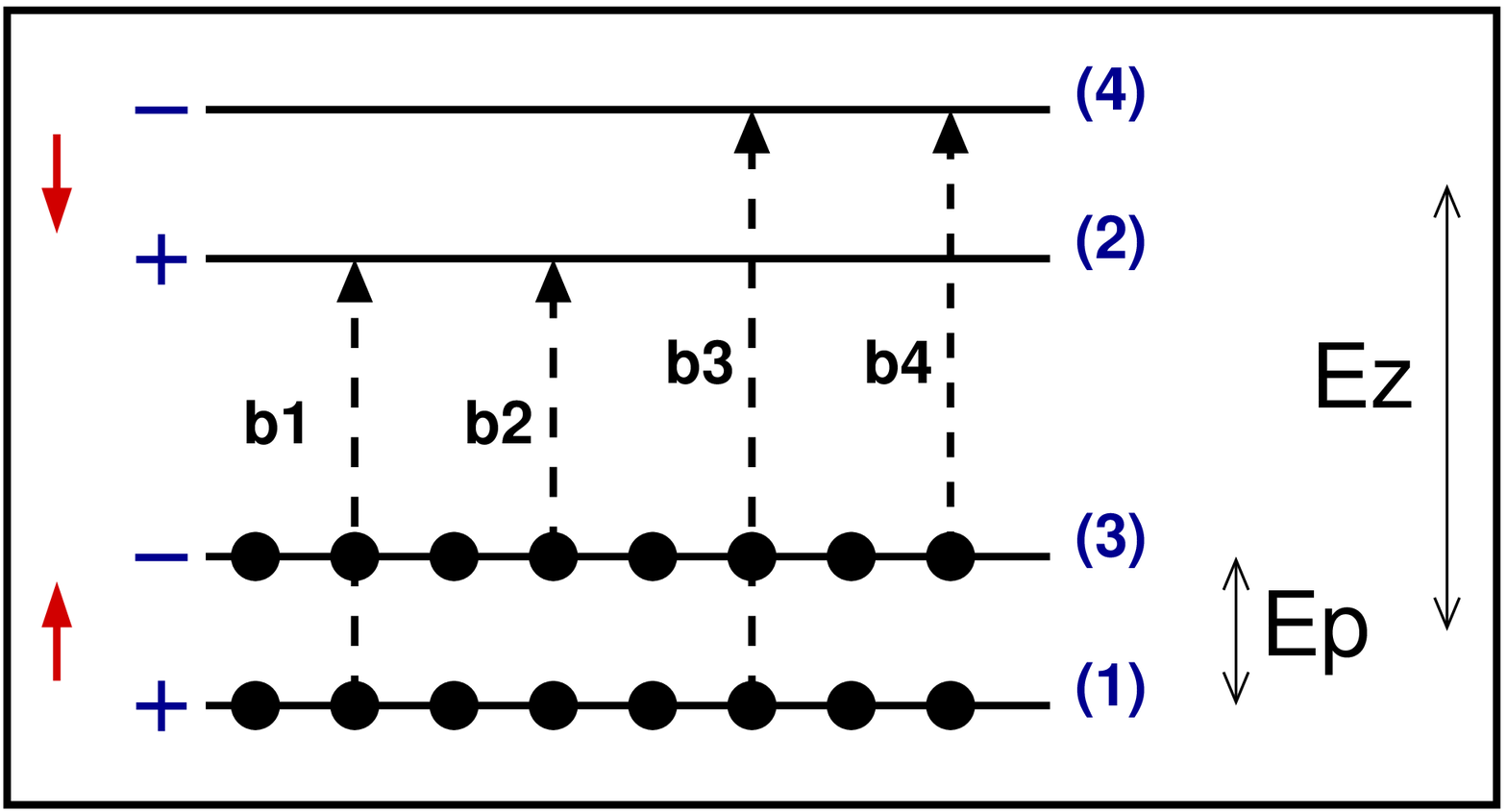}}
\caption{\label{fig02}{(color online) Schematic representation of the four highly
    degenerate lowest Landau levels when $\ez > \ep$. The state $|{\rm SFM} \rangle$
    is obtained by completely filling the energy levels (1) and (3). b1, b2, b3, and b4 are
    the elementary neutral excitations which are related to the
    density operators  $\rho_{21}(\bq)$, $\rho_{23}(\bq)$,
    $\rho_{41}(\bq)$, and $\rho_{43}(\bq)$ respectively.}}
\end{figure}

In this phase, $\ez > \ep$ and the total number of electrons ${\rm
N} = 2\nphi$. The ground state of the noninteracting Hamiltonian
(\ref{freehamiltonian2}) is a spin-polarized pseudospin-singlet
state,
\begin{equation}
|{\rm SFM} \rangle =
 \prod_{m=1}^{\nphi-1}c^\dagger_{m\,-\,\uparrow}c^\dagger_{m\,+\,\uparrow}|0\rangle.
\label{sfm}
\end{equation}
The particle-hole excitations are now created by the density
operators $\rho_{21}(\bq)$, $\rho_{23}(\bq)$, $\rho_{41}(\bq)$,
and $\rho_{43}(\bq)$ as it is illustrated in Fig. \ref{fig02}.

The commutation relations between $\rho_{IJ}(\bq)$ ($ I = 1,3$ and
$J = 2,4$) and their respective Hermitian conjugates
$\rho_{JI}(-\bq)$ read [see Eq. (\ref{commutator})]
\begin{eqnarray}
\nonumber
[\rho_{IJ}(\bq),\rho_{JI}(\bq')] &=&
e^{\bq\cdot\bq'l^2/2}
\left[e^{i\bq\wedge\bq'/2}\rho_{II}(\bq+\bq')\right.\\
\nonumber \\
&&-\left.
 e^{-i\bq\wedge\bq'/2}\rho_{JJ}(\bq+\bq')\right].
\label{commutator22}
\end{eqnarray}
Here, the expansions of $\rho_{11}(\bq)$, $\rho_{22}(\bq)$, $\rho_{33}(\bq)$,
and $\rho_{44}(\bq)$ around the reference state $|{\rm SFM} \rangle$
are given by
\begin{eqnarray}
\rho_{II}(\bq) &=& \langle {\rm SFM}|\rho_{II}(\bq)|{\rm SFM}\rangle +
\delta\rho_{II}(\bq)
\nonumber \\
\nonumber && \\
&=& \nphi(\delta_{I,1} + \delta_{I,3})\delta_{\bq,0} + \delta\rho_{II}(\bq),
\label{assumption11}
\end{eqnarray}
and therefore the commutation relations (\ref{commutator22})
reduce to (neglecting the density fluctuations
$\delta\rho_{II}(\bq)$)
\begin{equation}
\nonumber
[\rho_{IJ}(\bq),\rho_{JI}(\bq')] \approx
\delta_{\bq,-\bk}\nphi e^{(ql)^2/2}.
\label{commutator33}
\end{equation}
Using the same arguments of the previous section, we assume that
$\rho_{21}(\bq)$, $\rho_{41}(\bq)$, $\rho_{23}(\bq)$, and
$\rho_{43}(\bq)$ are approximately boson operators. Indeed, they
are independent operators because
\begin{eqnarray}
\rho_{IJ}(\bq) &=& \langle {\rm SFM}|\rho_{IJ}(\bq)|{\rm SFM}\rangle +
\delta\rho_{IJ}(\bq) \approx 0,
\label{assumption111}
\end{eqnarray}
for $(I,J) = (1,3)$ and $(4,2)$.

To sum up, the spin phase is characterized by a set of four independent
boson operators defined as
\begin{eqnarray}
b^\dagger_1(\bq)&\equiv&\alpha_q\rho_{21}(\bq),
         \;\;\;\;\;\;
b_1(\bq) \equiv \alpha_q\rho_{12}(-\bq),
\nonumber \\
b^\dagger_2(\bq)&\equiv&\alpha_q\rho_{23}(\bq),
         \;\;\;\;\;\;
b_2(\bq) \equiv \alpha_q\rho_{32}(-\bq),
\nonumber \\
b^\dagger_3(\bq)&\equiv&\alpha_q\rho_{41}(\bq),
         \;\;\;\;\;\;
b_3(\bq) \equiv \alpha_q\rho_{14}(-\bq),
\label{bosons2} \\
b^\dagger_4(\bq)&\equiv&\alpha_q\rho_{43}(\bq),
         \;\;\;\;\;\;
b_4(\bq) \equiv \alpha_q\rho_{34}(-\bq),
\nonumber
\end{eqnarray}
with $\alpha_q = e^{(lq)^2/4}/\sqrt{\nphi}$, and
obeying the canonical boson algebra (\ref{boson-algebra}).

The introduction of new boson operators implies that the
expansions of the density operators $\rho_{II}(\bq)$ in terms of
the bosons
are no longer given by Eqs. (\ref{boson-density1}) and
(\ref{boson-density2}). Following the same procedure discussed in the
previous section, it is possible to show that
\begin{eqnarray}
\nonumber \rho_{11}(\bq) &=& \nphi\delta_{\bq,0} -
                \sum_{\bk\; i=1,3}e^{-(lq)^2/4-i\bq\wedge\bk/2}
                b^\dagger_i(\bq+\bk)b_i(\bk), \\
\nonumber && \\
\nonumber \rho_{22}(\bq) &=&
                \sum_{\bk\; i=1,2}e^{-(lq)^2/4+i\bq\wedge\bk/2}
                b^\dagger_i(\bq+\bk)b_i(\bk), \\
\nonumber && \\
\nonumber \rho_{33}(\bq) &=& \nphi\delta_{\bq,0} -
                \sum_{\bk\; i=2,4}e^{-(lq)^2/4-i\bq\wedge\bk/2}
                b^\dagger_i(\bq+\bk)b_i(\bk), \\
\nonumber && \\
          \rho_{44}(\bq) &=&
                \sum_{\bk\; i=3,4}e^{-(lq)^2/4+i\bq\wedge\bk/2}
                b^\dagger_i(\bq+\bk)b_i(\bk).
\label{rhoii2}
\end{eqnarray}
By adding up the four terms above, one can see that the electron
density operator $\rho(\bq)$ [Eq. (\ref{totaldensity})] also has
the form (\ref{densityop-boson}) with the replacements
$\sum_{i=1}^3 \rightarrow \sum_{i=1}^4$ and $\nphi\delta_{\bq,0}
\rightarrow 2\nphi\delta_{\bq,0}$. However, the boson
representation of the z-components of the spin and pseudospin
density operators and the mixed operator $P_ZS_Z(\bq)$ are modified,
i.e.,
\begin{eqnarray}
\nonumber
S_Z(\bq) &=& \nphi\delta_{\bq,0} - e^{-(lq)^2/4} \\
\nonumber && \\
          &&  \times\sum_{i,\bk}
              \cos\left(\bq\wedge\bk/2\right)
              b^\dagger_i(\bq+\bk)b_i(\bk),
\label{spinop-boson2} \\
\nonumber && \\
I^\mu_Z(\bq) &=& \sum_{i,\bk}
              f^\mu_i(\bq,\bk)
              b^\dagger_i(\bq+\bk)b_i(\bk),
\label{ppinop-boson2}
\end{eqnarray}
where $\sum_i = \sum_{i=1}^4$, $I^\mu_Z(\bq) = \pz(\bq)$ and $\sz\pz(\bq)$, and the form
factors are given by
\begin{eqnarray}
\nonumber f^P_1(\bq,\bk) &=&
          - f^P_4(\bq,\bk) =
          ie^{-(lq)^2/4}\sin\left(\bq\wedge\bk\right/2), \\
          f^P_2(\bq,\bk) &=&
          - f^P_3(\bq,\bk) =
          e^{-(lq)^2/4}\cos\left(\bq\wedge\bk\right/2),
\label{formfactors-sfm}
\end{eqnarray}
and
\begin{eqnarray}
\nonumber f^{PS}_1(\bq,\bk) &=&
          - f^{PS}_4(\bq,\bk) =
          - e^{-(lq)^2/4}\cos\left(\bq\wedge\bk\right/2), \\
\nonumber f^{PS}_2(\bq,\bk) &=&
          - f^{PS}_3(\bq,\bk) =
          - ie^{-(lq)^2/4}\sin\left(\bq\wedge\bk\right/2).
\end{eqnarray}
Finally, the new bosonic expressions of $\rho_{12}(\bq)$,
$\rho_{21}(\bq)$, $\rho_{34}(\bq)$, and $\rho_{43}(\bq)$ are shown
in the Appendix \ref{appendix2} [see Eqs. (\ref{ops1-sfm}) and (\ref{ops2-sfm})].

\subsection{Pseudospin phase}
\label{pfm-phase}

\begin{figure}[t]
\centerline{\includegraphics[height=4.0cm]{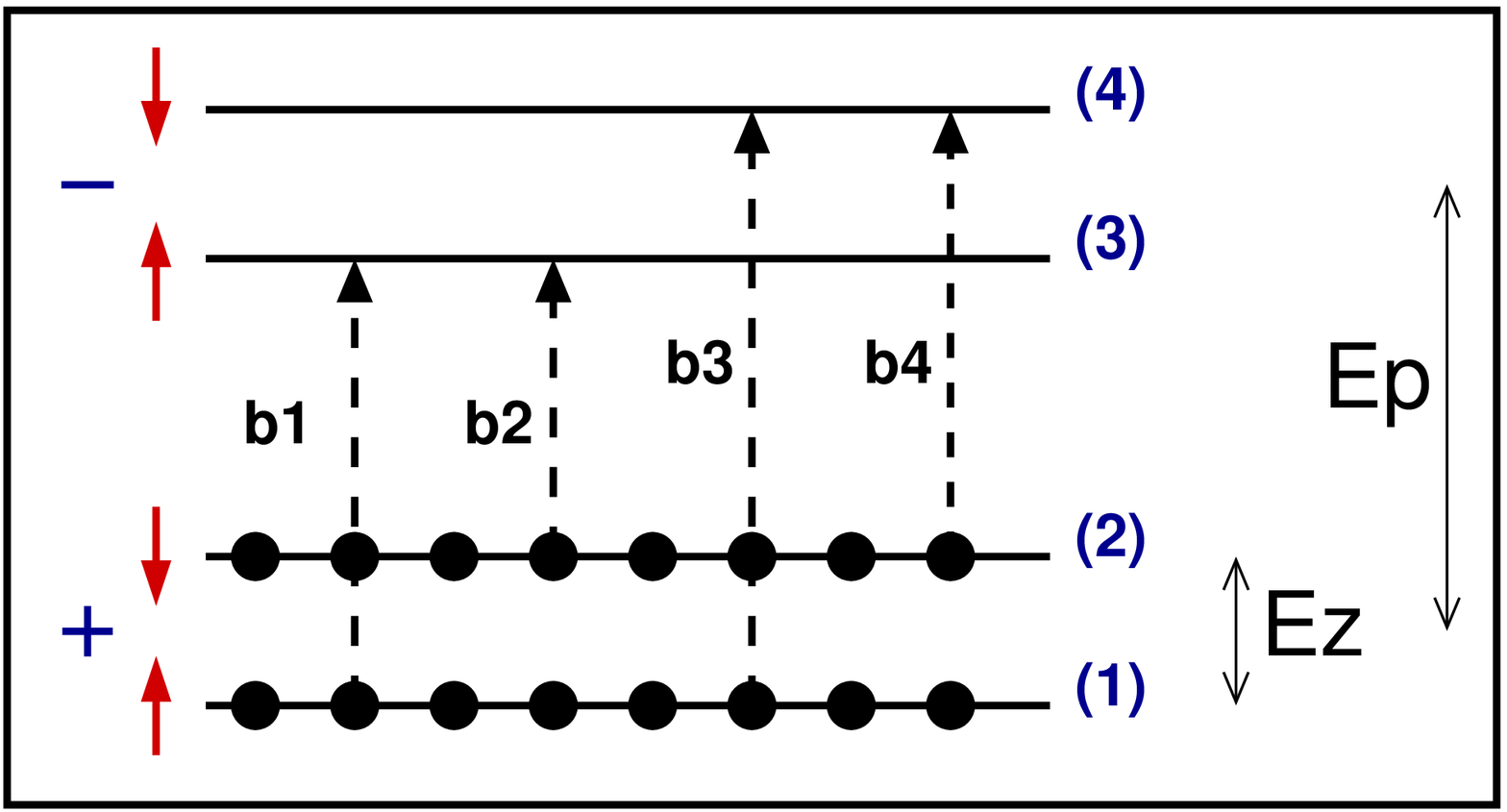}}
\caption{\label{fig03}{(color online) Schematic representation of the four highly
    degenerate lowest Landau levels when $\ez < \ep$. The state $|{\rm PFM} \rangle$
    is obtained by completely filling the energy levels (1) and (2). b1, b2, b3, and b4 are
    the elementary neutral excitations which are related to the
    density operators  $\rho_{31}(\bq)$, $\rho_{32}(\bq)$,
    $\rho_{41}(\bq)$, and $\rho_{42}(\bq)$ respectively.}}
\end{figure}

The situation here is quite similar to the one discussed in the
previous section, because again ${\rm N} =
2\nphi$ but now $\ez < \ep$. As a consequence,
the ground state of the noninteracting model
(\ref{freehamiltonian2}) is a spin-singlet pseudospin-polarized state,
\begin{equation}
|{\rm PFM} \rangle =
 \prod_{m=1}^{\nphi-1}c^\dagger_{m\,+\,\downarrow}c^\dagger_{m\,+\,\uparrow}|0\rangle,
\label{pfm}
\end{equation}
and the elementary neutral excitations are now related to
$\rho_{31}(\bq)$, $\rho_{32}(\bq)$, $\rho_{41}(\bq)$, and
$\rho_{42}(\bq)$ [see Fig. \ref{fig03}].

Again, one can show that the above four density operators give rise
to four independent boson operators, i.e.,
\begin{eqnarray}
b^\dagger_1(\bq)&\equiv&\alpha_q\rho_{31}(\bq),
         \;\;\;\;\;\;
b_1(\bq) \equiv \alpha_q\rho_{13}(-\bq),
\nonumber \\
b^\dagger_2(\bq)&\equiv&\alpha_q\rho_{32}(\bq),
         \;\;\;\;\;\;
b_2(\bq) \equiv \alpha_q\rho_{23}(-\bq),
\nonumber \\
b^\dagger_3(\bq)&\equiv&\alpha_q\rho_{41}(\bq),
         \;\;\;\;\;\;
b_3(\bq) \equiv \alpha_q\rho_{14}(-\bq),
\label{bosons3} \\
b^\dagger_4(\bq)&\equiv&\alpha_q\rho_{42}(\bq),
         \;\;\;\;\;\;
b_4(\bq) \equiv \alpha_q\rho_{24}(-\bq),
\nonumber
\end{eqnarray}
which satisfy the boson algebra (\ref{boson-algebra}). Indeed,
the commutator of each density operator with its corresponding
Hermitian conjugate is also given by Eq. (\ref{commutator22}) with
$I=1,2$ and $J=3,4$. The expansion (\ref{assumption11}) is
replaced by
\begin{eqnarray}
\rho_{II}(\bq) &=& \langle {\rm PFM}|\rho_{II}(\bq)|{\rm PFM}\rangle +
\delta\rho_{II}(\bq)
\nonumber \\
\nonumber && \\
&=& \nphi(\delta_{I,1} + \delta_{I,2})\delta_{\bq,0} + \delta\rho_{II}(\bq),
\end{eqnarray}
while Eq. (\ref{assumption111}) is preserved, but now  $(I,J) = (1,2)$
and $(4,3)$.

The set of creation and annihilation boson operators
(\ref{bosons3}) implies that Eqs. (\ref{rhoii2}) should be
replaced by
\begin{eqnarray}
\nonumber \rho_{11}(\bq) &=& \nphi\delta_{\bq,0} -
                \sum_{\bk,i=1,3}e^{-(lq)^2/4-i\bq\wedge\bk/2}
                b^\dagger_i(\bq+\bk)b_i(\bk), \\
\nonumber \rho_{22}(\bq) &=& \nphi\delta_{\bq,0} -
                \sum_{\bk,i=2,4}e^{-(lq)^2/4-i\bq\wedge\bk/2}
                b^\dagger_i(\bq+\bk)b_i(\bk), \\
\nonumber \rho_{33}(\bq) &=&
                \sum_{\bk,i=1,2}e^{-(lq)^2/4 + i\bq\wedge\bk/2}
                b^\dagger_i(\bq+\bk)b_i(\bk),\\
          \rho_{44}(\bq) &=&
                \sum_{\bk,i=3,4}e^{-(lq)^2/4+i\bq\wedge\bk/2}
                b^\dagger_i(\bq+\bk)b_i(\bk).
\label{rhoii3}
\end{eqnarray}
Again, the expression (\ref{densityop-boson}) for the electron
density operator is preserved, apart from the changes
$\sum_{i=1}^3 \rightarrow \sum_{i=1}^4$ and $\nphi\delta_{\bq,0}
\rightarrow 2\nphi\delta_{\bq,0}$. When compared with the results
of Sec.\ \ref{sfm-phase}, the boson representation of the
$z$-components of the spin, pseudospin, and mixed spin-pseudospin
density operators are interchanged, i.e.,
\begin{eqnarray}
I^\mu_Z(\bq) &=& \sum_{i,\bk}
              f^\mu_i(\bq,\bk)
              b^\dagger_i(\bq+\bk)b_i(\bk),
\label{spinop-boson3} \\
\nonumber && \\
\nonumber
P_Z(\bq) &=& \nphi\delta_{\bq,0} - e^{-(lq)^2/4} \\
          && \times\sum_{i,\bk}
             \cos\left(\bq\wedge\bk/2\right)
              b^\dagger_i(\bq+\bk)b_i(\bk),
\label{ppinop-boson3}
\end{eqnarray}
with $I^\mu_Z(\bq) = S_Z(\bq)$ and $P_ZS_Z(\bq)$, and
\begin{eqnarray}
\nonumber f^S_1(\bq,\bk) &=&
          - f^S_4(\bq,\bk) =
          ie^{-(lq)^2/4}\sin\left(\bq\wedge\bk\right/2), \\
\nonumber f^S_2(\bq,\bk) &=&
          - f^S_3(\bq,\bk) =
          e^{-(lq)^2/4}\cos\left(\bq\wedge\bk\right/2),
\end{eqnarray}
and
\begin{eqnarray}
\nonumber f^{PS}_1(\bq,\bk) &=&
          - f^{PS}_4(\bq,\bk) =
          - e^{-(lq)^2/4}\cos\left(\bq\wedge\bk\right/2), \\
\nonumber f^{PS}_2(\bq,\bk) &=&
          - f^{PS}_3(\bq,\bk) =
          - ie^{-(lq)^2/4}\sin\left(\bq\wedge\bk\right/2).
\end{eqnarray}
We again refer the reader to the Appendix \ref{appendix2} for the
boson representation of the operators $\rho_{12}(\bq)$,
$\rho_{21}(\bq)$, $\rho_{34}(\bq)$, and $\rho_{43}(\bq)$.

The generalization of the bosonization method\cite{doretto} for
the case of electrons restricted to the lowest Landau level and in
the presence of two discrete degrees of freedom is concluded.
The next sections will be devoted to an application of the
formalism.

\section{Quantum Hall ferromagnetism in graphene}

In this section, we apply the methodology developed above to study
the QHE at $\nu=\pm 1$ and $\nu = 0$ in graphene. We will follow the
lines of Ref.\ \onlinecite{doretto} and derive an effective boson
model for the system. Our starting point is the continuous model
for graphene recently proposed by Alicea and
Fisher.\cite{alicea06} Before outlining the derivation of this
model, we will briefly review some aspects of the Landau level
spectrum in graphene.

\subsection{Preliminaries on graphene}
\label{pre-graphene}

\begin{figure}[t]
\centerline{\includegraphics[height=7.0cm]{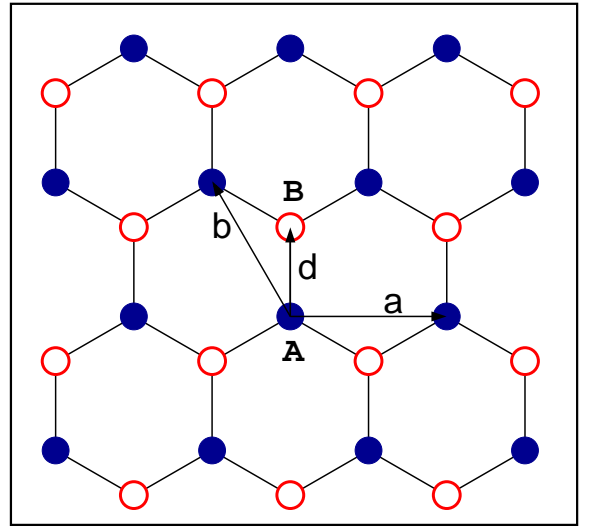}}
\caption{\label{fig04}{(color online) Schematic representation of
the honeycomb
    lattice. The triangular sublattices A and B are respectively
    represented by blue and red circles. $\mathbf{a}$ and
    $\mathbf{b}$ are the primitive vectors of the underline triangular
    sublattice A and $\mathbf{d}$ is the basis vector.}}
\end{figure}

Graphene is a collection of carbon atoms, which are arranged in a
two-dimensional honeycomb lattice, as it is
illustrated in Fig. \ref{fig04}.\cite{castroneto06,ando} The
lattice structure is triangular with two atoms per unit cell
located at the positions $(0,0)$ and $\mathbf{d} =
a_0(0,1/\sqrt{3})$. The lattice spacing is $a_0 = 2.46\;\AA$. It
might also be seem as two interpenetrating triangular sublattices A
and B. The primitive vectors of the (A) triangular lattice are
$\mathbf{a} = a_0(1,0)$ and $\mathbf{b} = a_0(-1/2,\sqrt{3}/2)$,
and therefore the primitive vectors of the reciprocal lattice are
$\mathbf{a}^* = (2\pi/a_0)(1,1/\sqrt{3})$ and $\mathbf{b}^* =
(2\pi/a_0)(0,2/\sqrt{3})$. In this atomic arrangement, the carbon
atoms are connected by strong covalent $\sigma$-bonds, derived from
the $sp^2$ hybridization of the atomic orbitals. The remaining
$p_z$ orbitals (perpendicular to the plane) have a weak overlap
and therefore they form a narrow band of $\pi$-orbitals, through
which the Fermi level passes.\cite{altland} By describing the
$\pi$-electrons within a tight-binding model
\begin{equation}
  \mathcal{H}_t = -t\sum_{\langle i,j \rangle}
                    \sum_{\sigma =\uparrow ,\downarrow}
                    \left( a^\dagger_{i,\sigma}b_{j\sigma} +
                    h.c.\right),
\label{tb-model}
\end{equation}
where $t \approx 2.7\,eV$ is the nearest-neighbor hopping energy
and the operators $a^\dagger_{i,\sigma}$ and
$b^\dagger_{i,\sigma}$ create a spin $\sigma$ electron on site $i$
of the sublattices A and B respectively, one can show that the
single-particle electron energy varies linearly with momentum
($\epsilon_q = \pm\hbar v_F|\bq|$, with $v_F = a_0\sqrt{3}t/2
\approx 10^6\,m/s$) around the six corners of the (hexagonal)
Brillouin zone, i.e., the band structure consists of six Dirac
cones. Only two of them are inequivalent, and here we consider the
ones around the points $\mathbf{K}=(2\pi/a_0)(2/3,0)$ and
$\mathbf{K}' = - (2\pi/a_0)(2/3,0)$. In the undoped case, there is
only one $\pi$-electron per carbon atom, the Fermi level lies at
the Dirac points and therefore the system is semi-metallic. By
using a gate-voltage, it is possible to modify the carriers,
either $p$-type or $n$-type (doped case).

The fact that the electronic structure of the system may be
described by an effective massless (continuous) Dirac model has
some important consequences. In particular, when a perpendicular
magnetic field is applied, a different Landau level structure
emerges when compared to the Schr\"odinger-like one observed in
the two-dimensional electron gas in GaAs heterostructures. Indeed,
one can show that the energy of the (Dirac) Landau levels are given by
\begin{equation}
E_{n,\sigma} = \mp \frac{1}{2}E_Z + {\rm sign(n)}\sqrt{2\hbar
v^2_F B |n|/c},
\label{dirac-ll}
\end{equation}
respectively for $\sigma =\,\uparrow$ and $\downarrow$, which
are associated with the two-components spinor eigenvectors
($n\not= 0$)
\begin{eqnarray}
|\hat{\Phi}_{n,m,\sigma,\alpha=+}\rangle &=& \frac{1}{\sqrt{2}}
   \left(\begin{array}{c}
           |n\, m\rangle \\
           {\rm sign(n)}|n - 1\, m\rangle \end{array} \right),
\nonumber \\
&& \label{ll-eigenstates1} \\
|\hat{\Phi}_{n,m,\sigma,\alpha=-}\rangle &=& \frac{1}{\sqrt{2}}
   \left(\begin{array}{c}
            {\rm sign(n)}|n-1\, m\rangle \\
            |n\, m\rangle\end{array} \right).
\nonumber
\end{eqnarray}
For $n=0$, we have
\begin{equation}
|\hat{\Phi}_{0,m,\sigma,+}\rangle =
   \left(\begin{array}{c}
           |0\, m\rangle  \\
           0\end{array} \right)
\;\;\;\;{\rm and}\;\;\;\; |\hat{\Phi}_{0,m,\sigma,-}\rangle =
   \left(\begin{array}{c}
           0 \\
           |0\, m\rangle \end{array} \right).
\label{ll-eigenstates2}
\end{equation}
Here, $\alpha = \pm$ corresponds, respectively, to $\mathbf{K}$ and
$\mathbf{K}'$ points, $m$ is the guiding center quantum number,
and $|n\, m\rangle$ are the Schr\"odinger's Landau level
eigenvectors. Each spinor component is related to one of the triangular
sublattices A and B. For $n\not= 0$, each eigenvector
$|\hat{\Phi}_{n, m,\sigma,+}\rangle$ and $|\hat{\Phi}_{n,
m,\sigma,-}\rangle$ has a weight (probability) equally distributed
between the two sublattices, while the lowest Landau level
eigenvectors $|\hat{\Phi}_{0,m,\sigma,+}\rangle$ and
$|\hat{\Phi}_{0,m,\sigma,-}\rangle$ are respectively localized on
sublattices A and B. The results
(\ref{dirac-ll})-(\ref{ll-eigenstates2}) show that the Dirac
Landau levels are approximately four-fold degenerate due to the electronic spin
and valley ($\alpha = \pm$) degrees of freedom.

Let us concentrate on the integer quantum Hall states in the
lowest Landau level ($n=0$). Apart from the fact that the fermion
field operator $\hat{\Psi}(\br)$ is a two-component spinor and the
momenta $\bq$ are measured with respect to the $\bkp$ and $\bkp'$
points [we refer the reader for a detailed discussion in the
Appendix \ref{appendix3}], the methodology developed in the
previous section can be used to study the QHE at $\nu=-1$ and
$\nu=0$. In fact, the former, which corresponds to a quarter
filled lowest Landau level, is associate with the spin-pseudospin
polarized phase (Sec.\ \ref{spfm-phase}), whereas the latter,
characterize by a half filled lowest Landau level, is associate
with either the spin (Sec.\ \ref{sfm-phase}) or the pseudospin
(Sec.\ \ref{pfm-phase}) phases.

\subsection{Alicea and Fisher's model}
\label{alicea-model}

The effective continuous model proposed by Alicea and Fisher to
study the quantum Hall effect in graphene goes beyond the
tight-binding approximation [see Ref.\ \onlinecite{alicea06} for
details]. In addition to $\mathcal{H}_t$ [Eq.(\ref{tb-model})], it
also includes the on-site electron-electron repulsion term
$\mathcal{H}_U$ and the (long range) Coulomb interaction
$\mathcal{H_{\rm Coul}}$, namely
\begin{equation}
  \mathcal{H} = \mathcal{H}_t + \mathcal{H}_U +
                \mathcal{H_{\rm Coul}},
\label{latticemodel}
\end{equation}
where
\begin{equation}
 \mathcal{H}_U = U\sum_i\left[\frac{1}{4}(\hat{n}_i)^2 -
                   \frac{1}{3}\mathbf{S}_i\cdot\mathbf{S}_i
                   \right]
\end{equation}
and
\begin{equation}
 \mathcal{H_{\rm Coul}} = \frac{1}{2}\sum_{i\not=j}
                       V(\br_i-\br_j)\hat{n}_i\hat{n}_j.
\end{equation}
Here, $U$ is the on-site repulsion energy and $V(r) = e^2/\epsilon r$
is the Coulomb potential, with an estimated dielectric constant
$\epsilon \approx 5$ [the energy scales for graphene are listed in the
Appendix \ref{appendix1}]. The electron number operator is $\hat{n}_i
= c^\dagger_{i,\uparrow}c_{i\uparrow} +
c^\dagger_{i,\downarrow}c_{i\downarrow}$, $\mathbf{S}_i =
(1/2)\sum_{\sigma,\lambda}c^\dagger_{i,\sigma}\hat{\sigma}_{\sigma\lambda}c_{i\lambda}$,
is the spin operator, where $\hat{\sigma}$ is a vector of Pauli
matrices, and  $c^\dagger_{i,\sigma} = a^\dagger_{i,\sigma}$ or
$b^\dagger_{i,\sigma}$ depending whether $i$ is on sublattice A or B.

Starting from the Hamiltonian (\ref{latticemodel}), a continuous
interacting theory was derived by expanding the fermion operators
$a^\dagger_{i,\sigma}$ and $b^\dagger_{i,\sigma}$ around the two
Dirac points $\mathbf{K}$ and $\mathbf{K}'$. After adding a
perpendicular magnetic field $\mathbf{B}=B\hat{z}$ and projecting
into the lowest Landau level, the model may be rewritten as
\begin{equation}
\mathcal{H} = \mathcal{H}_{SU4} + \mathcal{H}_{SB}
\label{continuous-model}
\end{equation}
where
\begin{equation}
\mathcal{H}_{SU4} = \frac{1}{2}\sum_\bq v(q)\rho(\bq)\rho(-\bq)
\label{hsu4}
\end{equation}
is the SU(4) invariant part of the Hamiltonian, with $v(q) = 2\pi
e^2/\epsilon q$ (the Fourier transform of the Coulomb potential in
two-dimensions), and\cite{nota}
\begin{eqnarray}
\nonumber \mathcal{H}_{SB} &=& -\ez\sz(\bq=0) -4\sum_\bq
v_1(\bq)\pz(\bq)\pz(-\bq)\\
\nonumber &&
+ u_0\sum_\bq \left[ \frac{1}{4}\rho(\bq)\rho(-\bq) +
\pz(\bq)\pz(-\bq) \right.\\
          &&
 \left. -\frac{1}{3}\mathbf{S}(\bq)\cdot\mathbf{S}(-\bq) -
\frac{1}{3}\pz\mathbf{S}(\bq)\cdot\pz\mathbf{S}(-\bq) \right]
\label{hsb}
\end{eqnarray}
contains terms that break the SU(4) symmetry. The parameter $u_0$ is
related to the on-site repulsion energy [$u_0 = \sqrt{3}a_0^2U/4$] and
$v_1(\bq)$ is the Fourier transform of
\begin{equation}
v_1(\br) = \frac{\sqrt{3}a_0^2}{8}\left[V(r + \hat{y}/\sqrt{3})
           - \left(1-\delta_{\br,0}\right)V(r)\right].
\label{u1-term}
\end{equation}
The model (\ref{continuous-model}) was analyzed in two distinct
situations: (i) the quantum Hall ferromagnetic regime, which
corresponds to an ideal, completely clean sample, and (ii) the
quantum Hall paramagnetic regime, where disorder effects are very
strong (very dirty sample). Here, we will only focus on the
quantum Hall ferromagnetic regime. The interplay between disorder
and electron-electron interactions will be postponed for a future
publication.

In order to derive an effective boson model for the quantum Hall
states at $\nu=-1$ and $\nu=0$, we just need to substitute the
respective boson representation of the electron density, the spin,
pseudospin and mixed spin-pseudospin density operators into the
Hamiltonian (\ref{continuous-model}) and normal order the
resulting expression. Although the expansion of the electron
density operator is similar for the three phases, each quantum
Hall state should be treated separately because the expansions in
terms of bosons of the spin/pseudospin density operators vary from
phase to phase.

\subsubsection{\bf Filling factor $\nu=-1$}

We start by considering the QHE at $\nu= -1$. The state at $\nu=
+1$ is related to it by particle-hole symmetry and will not be
discussed here.

{\it SU(4) invariant terms - }Let us firstly analyzed the SU(4)
invariant part of the Hamiltonian (\ref{continuous-model}).
Substituting the boson representation of the electron density operator
[Eq.(\ref{densityop-boson})] in $\mathcal{H}_{SU4}$ [Eq.(\ref{hsu4})]
and {\it normal ordering} the boson operators, apart from a constant
related to the positive background, we arrive at the following {\it
interacting} boson model
\begin{equation}
\mathcal{H}_{SU4}^B = \mathcal{H}_0^B + \mathcal{H}_I^B,
\label{boson-model}
\end{equation}
where the quadratic part is given by
\begin{equation}
\mathcal{H}_0^B = \sum_{i=1}^3\sum_\bq w_q b^\dagger_i(\bq)b_i(\bq)
\label{free-boson-model}
\end{equation}
and the quartic term reads
\begin{equation} \mathcal{H}_I^B =
  \sum_{i,j=1}^3\sum_{\bq,\bp,\bk}
   v_\bq(\bk,\bp)b^\dagger_i(\bk+\bq)b^\dagger_j(\bp-\bq)b_j(\bp)b_i(\bk).
\label{int-boson-model}
\end{equation}

The effective boson model (\ref{boson-model}) is the SU(4) counterpart
of the boson model derived in Ref.\ \onlinecite{doretto} for the
two-dimensional electron gas at $\nu=1$ realized in GaAs
heterostructures (hereafter called 2DEG at $\nu=1$). The ground state
of the model (\ref{boson-model}) is the boson vacuum, which is the
spin-pseudospin polarized state $|{\rm SPFM}\rangle$. Notice that this
state is indeed a spin polarized charge density wave (CDW) because the
electronic distribution is concentrated only in one sublattice [see
Eqs. (\ref{ll-eigenstates2}) and the discussion below this equation].
$\mathcal{H}^B_0$ describes three well-defined branches of bosonic
excitations, characterized by the same dispersion relation
\begin{equation}
w_q = \frac{e^2}{\epsilon l}\sqrt{\frac{\pi}{2}}\left[1 -
         e^{-(lq)^2/4}I_0\left((lq)^2/4\right)\right],
\label{su4-dispersion}
\end{equation}
where $I_0(x)$ is the modified Bessel function of the first
kind.\cite{arfken} Eq. (\ref{su4-dispersion}) is equal to the
dispersion relation of the elementary neutral excitations (magnetic
excitons) of the 2DEG at $\nu = 1$.\cite{doretto,kallin} In the long
wavelength limit, $w_q \approx \epsilon_B|l\bq|^2$ with $\epsilon_B =
\sqrt{\pi/32}\;(e^2/\epsilon l)$, and therefore the branch $i=1$
corresponds to spin wave excitations, while the branches $i=2$ and
$i=3$ to pseudospin wave and mixed spin-pseudospin wave excitations,
respectively [see Fig. \ref{fig01}]. At short wavelengths $w_q \approx
\sqrt{\pi/2}\;(e^2/\epsilon l)$, which is the energy of a very-well
separated particle-hole pair.\cite{kallin} Finally, the boson-boson
interaction potential is given by
\begin{equation}
v_\bq(\bk,\bp) =
2v(q)e^{-(lq)^2/2}\sin\left(\bq\wedge\bk\right/2)\sin\left(\bq\wedge\bp\right/2).
\label{su4-interaction}
\end{equation}
Apart from the fact that Eq. (\ref{int-boson-model}) describes
scattering processes between bosons within the same ($i=j$) and
different ($i\not=j$) branches, the interaction potential
(\ref{su4-interaction}) is similar to the one derived in Ref.\
\onlinecite{doretto}. It is worth mentioning that our approach
also provides an interaction between the bosonic excitations, which is
not captured by the analysis presented in Ref.\
\onlinecite{alicea06}.

Due to the similarities between the quantum Hall system in
graphene at $\nu=-1$ and the 2DEG at $\nu=1$, we would expect that
the charged excitations of the former might be described by
topological solitons\cite{sondhi,sondhi99} (quantum Hall skyrmions) as
well. In fact, the situation here is formally identical to the one
in the (spinfull) bilayer QHS at $\nu_T = 1$ in GaAs
heterostructures. The similarity clearly appears when the
upper-layer and down-layer electronic states are combined into the
bounding and anti-bounding states. In this case, there are four
possible kinds of charged excitations with topological charge $Q_T
= \pm 1$ (and corresponding electric charge $Q_e = eQ_T$), namely
one skyrmion ($Q_T = 1$ and $Q_e = e$) and three types of
antiskyrmions ($Q_T = -1$ and $Q_e = -e$). Indeed, they might be
considered as SU(4) skyrmions because the topological excitation
created by introducing an extra electric charge should involve the
three branches of neutral excitations in order to minimize the
total energy [see Ref.\  \onlinecite{ezawa04} for a detailed
description of SU(4) skyrmions in the context of the bilayers].

For the SU(2) version of the model (\ref{boson-model}), we know that
the boson-boson interaction potential (\ref{su4-interaction}) gives
rise to bound states of two-bosons which are related to small [SU(2)]
skyrmion-antiskyrmion pair excitations.\cite{doretto} The fact that
(\ref{su4-interaction}) describes scattering processes between
different bosonic branches indicates that here we would expect bound
states constituted by bosons belonging to the same and distinct
branches. Moreover, it was also shown that by describing the
topological excitation as a coherent state of bosons $|{\rm
sk}\rangle$ [see Eq. (64) of Ref. \onlinecite{doretto}], the
expectation value of the SU(2) boson model with respect to the state
$|{\rm sk}\rangle$ is equal to the energy functional derived from the
phenomenological theory of Sondhi {\it et al.}\cite{sondhi} for the
quantum Hall skyrmion, i.e., the {\it semiclassical limit} of the
SU(2) boson model agrees with Sondhi's theory for the quantum Hall
skyrmion.

It is quite straightforward to derive the semiclassical limit of the
interacting boson model (\ref{boson-model}). We start by writing down
the SU(4) counterpart of the state $|{\rm sk}\rangle$,
\begin{equation}
|{\rm sk}\rangle = \exp\left(\mathcal{B}\sum_{i=1}^3\sum_{\bq}
                       \Omega^i_\bq b^\dagger_i(-\bq) +
                       \bar{\Omega}^i_\bq b_i(\bq)
                    \right)|{\rm SPFM}\rangle,
\label{skyrmion}
\end{equation}
where $(\bar{\Omega}^i_{-\bq})^* = \Omega^i_\bq$ and the constant
$\mathcal{B}$ will be determined latter. With the aid of the
Baker-Hausdorff formula, one can show that the expectation value in
the state $|{\rm sk}\rangle$ of normal ordered boson operators is
obtained just by replacing each $b^\dagger_i(\bq)$ and $b_i(\bq)$
respectively for $i\mathcal{B}\bar{\Omega}^i_\bq$ and
$-i\mathcal{B}\Omega^i_{-\bq}$. Defining the excess charge
$\delta\rho(\bq) = \langle sk | \rho(\bq) | sk \rangle -
\nphi\delta_{\bq,0}$, where $\rho(\bq)$ is the electron density
operator (\ref{densityop-boson}), we have
\begin{equation}
\delta\rho(\bq) = 2i\sum_{i=1}^3\sum_\bk
        e^{-(lq)^2/2}\sin\left(\bq\wedge\bk/2\right)\bar{\Omega}^i_{\bk+\bq}\Omega^i_\bk.
\label{carga-top1}
\end{equation}
Assuming that the Fourier transform of $\bar{\Omega}^i_\bq$ and
$\Omega^i_\bq$ vary slowly in space, we can restrict ourselves to the
long wavelength limit of Eq. (\ref{carga-top1}), i.e., we can consider
$e^{-(lq)^2/2}\sin\left(\bq\wedge\bk/2\right) \approx \bq\wedge\bk/2$.
Within this approximation, the Fourier transform of $\delta\rho(\bq)$
is given by
\begin{equation}
\delta\rho(\br) =
     i\mathcal{B}^2l^2\sum_i\hat{z}\cdot\nabla\bar{\Omega}^i(\br)
    \times\nabla\Omega^i(\br),
\label{carga-top}
\end{equation}
which is in agreement with the expression for the topological charge
density derived by Arovas {\it et al.}\cite{arovas99} in their studies
of SU(N) quantum Hall skyrmions. Indeed, by comparing Eq.
(\ref{carga-top}) with Eq. (3) from Ref. \onlinecite{arovas99}, one
concludes that $\mathcal{B} = 1/\sqrt{2\pi l^2}$. Once the constant
$\mathcal{B}$ is fixed, we can now calculate $\langle sk|
\mathcal{H}^B_{SU4}|sk\rangle$ and show that
\begin{eqnarray}
\langle \mathcal{H}^B_{SU4} \rangle &=&
                      2\rho^0_S\sum_i\int\;d^2r|\nabla\Omega^i(\br)|^2
\nonumber \\
&+& \frac{1}{2}\int\;d^2rd^2r'v\left(|\br-\br'|\right)
    \delta\rho(\br)\delta\rho(\br'),
    \;\;\;\;
\label{sondhi-models}
\end{eqnarray}
where $\rho^0_S = 1/(16\sqrt{2\pi})\;e^2/\epsilon l$ is the stiffness
and $v(r)= e^2/\epsilon r$ is the Coulomb potential. The energy
functional $E[\Omega^i(\br)] = \langle \mathcal{H}^B_{SU4} \rangle$
[Eq. (\ref{sondhi-models})], which corresponds to the SU(4)
counterpart of Sondhi's model, agrees with the findings of Arovas and
co-workers.\cite{arovas99} This analysis shows that the boson model
(\ref{boson-model}) can indeed be used to study SU(4) quantum Hall
skyrmions in graphene.

{\it Symmetry breaking terms} - The degeneracy of the three branches
of boson excitations is lifted when the SU(4) symmetry breaking term
$\mathcal{H_{SB}}$ is taken into account. Following the same procedure
used above, we can derive an effective boson model from the
Hamiltonian (\ref{hsb}). The task here is slightly more difficult
because $\mathcal{H_{SB}}$ involves more complex expressions.

Let us start by expanding the operators $\mathbf{S}(\bq)$ and
$\pz\mathbf{S}(\bq)$ in terms of the density operators
$\rho_{IJ}(\bq)$. It is possible to show that
\begin{eqnarray}
\nonumber
\mathbf{S}(\bq)\cdot\mathbf{S}(-\bq) &+&
\pz\mathbf{S}(\bq)\cdot\pz\mathbf{S}(-\bq) =   \\
\nonumber && \\
\nonumber \sz(\bq)\sz(-\bq) &+& \pz\sz(\bq)\pz\sz(-\bq) +
\rho_{12}(\bq)\rho_{21}(-\bq)  \\
\nonumber && \\
\nonumber
+ \, \rho_{21}(\bq)\rho_{12}(-\bq) &+& \rho_{34}(\bq)\rho_{43}(-\bq) +
\rho_{43}(\bq)\rho_{34}(-\bq).
\end{eqnarray}
The boson representations of the above density operators
$\rho_{IJ}(\bq)$ are shown in the Appendix \ref{appendix2} [see Eqs.\
  (\ref{ops1-spfm}) and (\ref{ops2-spfm})].
After a lengthy but straightforward calculation, one can show that
$\mathcal{H}_{SB}$ is also mapped into an interacting boson
model. Adding Eq. (\ref{boson-model}), which was derived from the
SU(4) invariant term, the total effective boson model may be written as
\begin{equation}
\mathcal{H}^B = \bar{\mathcal{H}}_0^B + \bar{\mathcal{H}}_I^B.
\label{total-boson-model1}
\end{equation}
The quadratic term now reads
\begin{equation}
\bar{\mathcal{H}}_0^B = \sum_{i=1}^3\sum_\bq\bar{\omega}_i(\bq)
b^\dagger_i(\bq)b_i(\bq), \label{total-boson-model1.0}
\end{equation}
where $\bar{\omega}_i(q)$ are the renormalized boson dispersion
relations,
\begin{eqnarray}
\nonumber \bar{\omega}_1(q) &=& \ez + 2(u_0-u_1)\nphi\left(1 -
e^{-(lq)^2/2}\right) +
                 w_q, \\
\nonumber \bar{\omega}_2(q) &=& 4u_1\nphi
                 -4\sum_\bk
                 v_1(\bk)e^{-(lk)^2/2}\\
\nonumber
                 && \times\cos^2\left(\bk\wedge\bq/2\right)
                 + w_q, \\
           \bar{\omega}_3(q) &=& \ez + \bar{\omega}_2(q),
\label{dispersion-nu1}
\end{eqnarray}
with $w_q$ given by Eq. (\ref{su4-dispersion}) and $u_1 = v_1(\bq =
0)$ [see Eq. (\ref{u1-term})]. In the small momentum region, we have
\begin{equation}
\bar{\omega}_1(q) \approx \Delta_i + 4\pi\rho^i_S|l\bq|^2,
\label{approx-dispersion}
\end{equation}
where the excitation gaps $\Delta_i$ and the renormalized stiffnesses
are given by
\begin{eqnarray}
\nonumber \Delta_1 &=& \ez,
             \;\;\;\;\;\;\;\;\;\;\;\; \Delta_2 = \sqrt{\pi^3/24}(a_0/l)u_1\nphi, \\
\nonumber
\Delta_3 &=& \Delta_1 + \Delta_2, \\
\nonumber \rho^1_S &=& \nphi(u_0-u_1)/4\pi + \rho^0_S, \;\;\;\;\;\;
\\
\rho^2_S &=& \rho^3_S = u_1\nphi/4\pi + \rho^0_S.
\label{approx-dispersion-spfm}
\end{eqnarray}
Notice that $\bar{\omega}_2(q = 0) \ll \bar{\omega}_1(q = 0)$ and
$\bar{\omega}_3(q = 0)$. Both small and large momentum limits of Eqs.
(\ref{dispersion-nu1}) agree with the results derived by Alicea and
Fisher.\cite{alicea06} The interaction term assumes the form
\begin{equation}
\bar{\mathcal{H}}_I^B =
\sum_{i,j=1}^3\sum_{\bq,\bp,\bk}\bar{v}^{i,j}_\bq(\bk,\bp)b^\dagger_i(\bk+\bq)
             b^\dagger_j(\bp-\bq)b_j(\bp)b_i(\bk),
\label{total-boson-model1.1}
\end{equation}
where the total boson-boson interaction potential
$\bar{v}^{i,j}_\bq(\bk,\bp)$, which is richer than the one derived
only from the SU(4) invariant part of the Hamiltonian
(\ref{continuous-model}), is given by
\begin{eqnarray}
\nonumber
 \bar{v}^{i,j}_\bq(\bk,\bp) &=&
    \delta_{i,j}\frac{2u_0}{3}e^{-(lq)^2/2}
    \left(\sin\left(\bq\wedge\bk/2\right)\sin\left(\bq\wedge\bp/2\right)\right. \\
\nonumber \\ \nonumber
          &-& \left. \cos\left(\bq\wedge\bk/2\right)\cos\left(\bq\wedge\bp/2\right)
          \right)  \\
\nonumber \\ \nonumber
     &+&   \delta_{i,1}\delta_{j,1}\frac{2u_0}{3}
            e^{-l^2|\bq + \bk|^2/2}\cos\left(\bq\wedge(\bk - \bp)/2\right) \\
\nonumber \\ \nonumber
     &+&   4\left(\frac{u_0}{3}  - u_1\right)f^P_i(\bq,\bk)f^P_j(-\bq,\bp)\\
\nonumber \\ \nonumber
      &+& \bar{\delta}_{i,j}\frac{u_0}{3}e^{-(lq)^2/2}\left(
           2\sin\left(\bq\wedge\bk/2\right)\sin\left(\bq\wedge\bp/2\right)\right.\\
\nonumber \\ \nonumber
      &+& \left. e^{i\bq\wedge(\bk - \bp)/2}   \right) \\
\nonumber \\ \nonumber
      &+& \frac{2u_0}{3}e^{-i\bq\wedge(\bk - \bp)/2}\left(
           \delta_{i,1}(1-\delta_{j,1})e^{-l^2|\bq + \bk|^2/2}\right.\\
\nonumber \\
      &-& \left. \delta_{i,2}\delta_{j,3}e^{-l^2|\bq + \bk -
           \bp|^2/2}\right),
\label{potential-spfm}
\end{eqnarray}
with the form factors $f^P_i(\bq,\bk)$ given by Eqs.
(\ref{formfactors-spfm}) and $\bar{\delta}_{i,j} = 1 - \delta_{i,j}$.
Finally, it is worth mentioning that the ground state of the system is
still the boson vacuum $|{\rm SPFM}\rangle$. Indeed, this result is
corroborated by exact diagonalizations on small systems.\cite{sheng07}

The introduction of new terms in the boson-boson interaction potential
might modify the two-bosons spectrum, for instance, one particular
kind of bound state may have lower energy than the others. As a
consequence, one specific type of skyrmion-antiskyrmion pair
excitation will be more favorable. Indeed, it was argued that a
pseudospin skyrmion-antiskyrmion excitation should determine the
charge gap due to the smallness of the excitation gap
$\bar{\omega}_2(q = 0)$.\cite{alicea06}

Although  $\bar{v}^{i,j}_\bq(\bk,\bp)$ is quite complex, it is
possible to make a simple analysis by considering the SU(2) limit of
the bosonic Hamiltonian (\ref{total-boson-model1}) and then by
calculating the semiclassical limit of this reduced boson model. In
this case, Eq. (\ref{skyrmion}) simplifies to $ |{\rm sk}\rangle =
\exp(\mathcal{B}\sum_{\bq} \Omega^i_\bq b^\dagger_i(-\bq) +
\bar{\Omega}^i_\bq b_i(\bq))|{\rm SPFM}\rangle, $ i.e., we assume that
only the $i-th$ bosonic branch is excitated while the others are kept
frozen. Following the same steps which lead to Eq.
(\ref{sondhi-models}) and approximating $e^{-(lq)^2/2}
\cos\left(\bq\wedge\bk/2\right)\cos\left(\bq\wedge\bp/2\right) \approx
e^{-(lq)^2/2}$, one can show that the functional energy for the
different skyrmion flavors assumes the form
\begin{eqnarray}
\nonumber
E_i[\mathbf{n}(\br)] &=& E_i^\Delta[\mathbf{n}] + E_i^G[\mathbf{n}]
                         + E_i^{ZZ}[\mathbf{n}] + E_i^C[\mathbf{n}] \\
\nonumber \\ \nonumber
  &=& \int\;d^2r \left[ 2\mathcal{B}^2(1-\Delta_i n^z(\br))
             +  2\rho^i_S(\nabla\mathbf{n}(\br))^2 \right]\\
\nonumber \\ \nonumber
  &+& \frac{1}{2}\int\;d^2rd^2r'\mathcal{B}^4\tilde{u}_ie^{-|\br-\br'|^2/2l^2} \\
\nonumber
      && \times\left(1-n^z(\br)\right)\left(1-n^z(\br')\right)\\
\nonumber \\ \nonumber
  &+& \frac{1}{2}\int\;d^2rd^2r'\left(\frac{e^2}{\epsilon|\br-\br'|} +
      \tilde{v}_i e^{-|\br-\br'|^2/2l^2}\right) \\
      && \times\delta\rho(\br)\delta\rho(\br'),
      \;\;\;\;
\label{sondhi-models1}
\end{eqnarray}
where $i=1,2,3$ refer respectively to spin, pseudospin, and mixed
spin-pseudospin-like skyrmions. $\mathbf{n}(\br)$ is a unit vector
defined by the relation $\Omega(\br) = \hat{z}\times\mathbf{n}(\br)$
[see Ref. \onlinecite{doretto} for details]. $\Delta_i$ and $\rho^i_S$
are given by Eqs. (\ref{approx-dispersion-spfm}), $\tilde{u}_1 = 0$,
$\tilde{u}_2 = \tilde{u}_3 = 32\nphi(u_0/6 - u_1)$, $\tilde{v}_1 =
2\nphi(2u_0/3 - u_1)$, and $\tilde{v}_2 = \tilde{v}_3 = \nphi u_0/3$.

Notice that the SU(4) symmetry breaking part of the Hamiltonian
(\ref{continuous-model}) adds to the energy functional for the
skyrmion a Zeeman-like term ($E_i^\Delta[\mathbf{n}]$) and provides
small contributions to both the stiffness and the
topological-charge-topological-charge interaction potential. In fact,
$\tilde{v}_2$ and $\tilde{v}_3 > 0$ while $\tilde{v}_1$ can be either
positive or negative depending on the value of the on-site repulsion
energy $U$. For the pseudospin and the mixed spin-pseudospin-like
skyrmions (boson branches $i=2$ and $3$, respectively) there is an
extra contribution given by $E_i^{ZZ}[\mathbf{n}]$. This term favors
excitations with $n^z(\br) \leq 0$ because $4\nphi(u_0/6 - u_1) < 0$
when $2 < U < 12\, eV$ [see Appendix \ref{appendix1}]. Remembering
that a quantum Hall skyrmion is characterized by $n^z(\br) \rightarrow
1$ when $r \rightarrow \infty$, one might conclude that
$E_i^{ZZ}[\mathbf{n}]$ contributes to an increase of the radius of the
skyrmion and so its stability.

It is still difficult to predict which type of skyrmion has the lowest
energy without performing careful calculations. What we can easily see
is that if the on-site repulsion energy is $U \approx 10\;eV$, then
the scenario proposed in Ref. \onlinecite{alicea06}, that a pseudospin
skyrmion should be the lowest energy one, is confirmed. In this case
$\rho^1_S \approx \rho^2_S \approx \rho^3_S$ and $\tilde{v}_1 \approx
\tilde{v}_2 \approx \tilde{v}_3 \ll \epsilon_c$. As the skyrmion
energy is $4\pi\rho^i_S +
\mathcal{O}(\Delta_i/\epsilon_C)$,\cite{sondhi,sondhi99} the lowest
energy soliton should be the one related with the excitation branch
which has the smallest $\Delta_i$, i.e., the pseudospin branch
($i=2$). Notice that the presence of $E_i^{ZZ}[\mathbf{n}]$ for $i=2$
and $3$ does not alter the above conclusions because this term should
reduce the total energy.

Finally, it is worth mentioning that (\ref{potential-spfm}) contains
$\cos\left(\bq\wedge\bk/2\right)\cos\left(\bq\wedge\bp/2\right)$ like
terms, which are also present in the boson-boson interaction potential
derived for the bilayer QHS at $\nu_T=1$ (spinless case) within the
SU(2) bosonization method.\cite{doretto06} Such similarity implies
that, in principle, a Bose-Einstein condensate could be realized here.
Let us consider again the SU(2) limit of the boson model
(\ref{total-boson-model1}) and focus, for instance, on the mixed
spin-pseudospin branch ($i=3$). The phase with $\nphi/2$ bosons should
then correspond to the antiferromagnetic one proposed by
Herbut.\cite{herbut07} Assuming that the bosons condense in their
lowest energy mode ($\bq=0$) and treating the reduced boson model
within the Bogoliubov approximation, one arrives at a model similar to
Eq. (8) from Ref. \onlinecite{doretto06} with the replacement
$\lambda_\bq \rightarrow 8(u_0/6 - u_1)\exp(-(lq)^2/2)$. As it was
showed in the last paragraph, $u_0/6 - u_1 < 0$, and therefore such a
phase should be unstable. Similar considerations hold for the
pseudospin branch. The situation is more delicate for the spin wave
branch and it will not be discussed here.

\subsubsection{\bf Filling factor $\nu=0$}

The analysis of the quantum Hall state at $\nu=0$ follows the same
lines of the previous section with the difference that now either the spin or
the pseudospin phases can be realized.

{\it Spin phase -} Let us firstly assume that the system is in the
spin phase. In this case, an effective boson model can be obtained
from the fermionic Hamiltonian (\ref{continuous-model}) with the aid
of the expressions calculated in Sec. \ref{sfm-phase} and Eqs.
(\ref{ops1-sfm}) and (\ref{ops2-sfm}). Due to the fact that the boson
representation of the electron density operator (\ref{totaldensity})
does not change from phase to phase, the boson model derived from the
SU(4) invariant part of the total Hamiltonian [Eq. (\ref{hsu4})] is
similar to Eq. (\ref{boson-model}) with the replacement $\sum_{i=1}^3
\rightarrow \sum_{i=1}^4$. The ground state of the system is also the
boson vacuum, which now corresponds to the state $|{\rm SFM}\rangle$.
There are four branches of well-defined bosonic excitations. In the
small momentum region, the branches $i=1$ and $4$ describe spin wave
excitations whereas the branches $i=2$ and $3$ correspond to mixed
spin-pseudospin wave excitations [see Fig. \ref{fig02}].

The total effective boson model, which also includes the terms obtained
from $\mathcal{H}_{SB}$, reads
\begin{equation}
\mathcal{H}^B = \bar{\mathcal{H}}_0^B + \bar{\mathcal{H}}_I^B +
                \mathcal{V}^B.
\label{total-boson-model2}
\end{equation}
The quadratic term $\bar{\mathcal{H}}_0^B$ is again given by
Eq. (\ref{total-boson-model1.0}) with 
%
\begin{eqnarray}
\nonumber
  \bar{\omega}_1(q) &=& \bar{\omega}_4(q) \\
\label{dispersion-nu01}
   &=& \ez + 2(u_0-u_1)\nphi\left(1 - e^{-(lq)^2/2}\right) + w_q, \\
\nonumber
 \bar{\omega}_2(q) &=& \bar{\omega}_3(q)  \\
\nonumber
 &=& \ez + 2u_0\nphi - 2u_1\nphi\left(1 + e^{-(lq)^2/2}\right) +
                 w_q,
\end{eqnarray}
where $w_q$ is given by Eq. (\ref{su4-dispersion}). In the small
momentum region, $\bar{\omega}_i(q)$ assume the form
(\ref{approx-dispersion}), with the following excitation gaps
$\Delta_i$ and stiffness $\rho^i_S$
\begin{eqnarray}
\nonumber
\Delta_1 &=& \Delta_4 = \ez, \;\;\;\;\;\;\;\;\;  \\
\nonumber
\Delta_2 &=& \Delta_3 = \ez + 2\nphi(u_0 - 2u_1),\\
&& \label{approx-dispersion-sfm}  \\
\nonumber
\rho^1_S &=& \rho^4_S =  \nphi(u_0 - u_1)/4\pi + \rho^0_S, \\
\nonumber
\rho^2_S &=& \rho^3_S =  \nphi u_1/4\pi  + \rho^0_S.
\end{eqnarray}
Notice that the introduction of the symmetry breaking terms does not
modify the ground state of the system $|{\rm SFM}\rangle$.

The boson-boson interaction part of the total Hamiltonian has two
distinct terms. The first one, $\bar{\mathcal{H}}_I^B$, is equal to
Eq. (\ref{total-boson-model1.1}), but now the interaction potential
$\bar{v}^{i,j}_\bq(\bk,\bp)$ reads
\begin{eqnarray}
\nonumber
 \bar{v}^{i,j}_\bq(\bk,\bp) &=& u_0e^{-(lq)^2/2}
    \left(\sin\left(\bq\wedge\bk/2\right)\sin\left(\bq\wedge\bp/2\right)\right.\\
\nonumber \\ \nonumber
         &-& \left.\frac{1}{3}\cos\left(\bq\wedge\bk/2\right)\cos\left(\bq\wedge\bp/2\right)
          \right) \\
\nonumber \\ \nonumber
          &-& \delta_{i,j}\frac{u_0}{3}\left(e^{-(lq)^2/2} -
              2(\delta_{i,1}+\delta_{i,4})\right.
\\ \nonumber \\ \nonumber
           &&\left. \times e^{-l^2|\bq + \bk|^2/2}\right)\cos\left(\bq\wedge(\bk - \bp)/2\right)
\\ \nonumber \\ \nonumber
           &+& 4\left(\frac{u_0}{3}  - u_1\right)f^P_i(\bq,\bk)f^P_j(-\bq,\bp)\\
\nonumber \\ \nonumber
          &+& \bar{\delta}_{i,j}\frac{u_0}{3}\left[
               h_{i,j}e^{-(lq)^2/2}\cos\left(\bq\wedge(\bk - \bp)/2\right)
               \right.\\
\nonumber \\ \nonumber
          &+& \left. \bar{h}_{i,j}\left(ie^{-(lq)^2/2}\sin\left(\bq\wedge(\bk -
          \bp)/2\right)\right. \right.
\\ \nonumber \\
              &+& \left.\left.  e^{-l^2|\bq + \bk|^2/2}e^{-i(-1)^{i+j}\bq\wedge(\bk -
              \bp)/2}\right)
               \right],
\end{eqnarray}
with the form factors $f^P_i(\bq,\bk)$ given by Eqs.
(\ref{formfactors-sfm}),
$h_{i,j} = \delta_{i,1}\delta_{j,4} + \delta_{i,4}\delta_{j,1} +
\delta_{i,2}\delta_{j,3} + \delta_{i,3}\delta_{j,2}$, and
$\bar{h}_{i,j} = 1 - h_{i,j}$ The second one, $\mathcal{V}^B$, can be
written as
\begin{equation}
\mathcal{V}^B =
\sum_{\bq,\bp,\bk}v'_\bq(\bk,\bp)b^\dagger_1(\bk+\bq)b^\dagger_4(\bp-\bq)b_3(\bp)b_2(\bk),
\label{int-boson-model2}
\end{equation}
where
\begin{eqnarray}
\nonumber
v'_\bq(\bk,\bp) &=& \frac{2u_0}{3}\exp(i\bq\wedge(\bp-\bk)/2) \\
                       &\times& \left(e^{(-|l(\bk+\bq)|^2} + e^{-|l(\bp-\bq)|^2}\right).\;\;\;\;
\label{coef-int-boson-model2}
\end{eqnarray}

{\it Pseudospin phase-} Turning to the pseudospin phase, similar
considerations show that this phase is also characterized by an
effective boson model analogous to (\ref{total-boson-model2}). The
ground state is the boson vacuum $|{\rm PFM}\rangle$ [Eq. (\ref{pfm})]
and the dispersion relations of the four branches of bosonic
excitations are
\begin{eqnarray}
\nonumber
   \bar{\omega}_i(q) &=& \ez(\delta_{i,3} - \delta_{i,2}) - 2u_0\nphi
\\
                     &+& 2u_1\nphi\left(3 - e^{-(lq)^2/2}\right) + w_q,
\label{dispersion-nu02}
\end{eqnarray}
for $i=1,\,2,\,3$ and $4$. The long wavelength limit behavior of
$\bar{\omega}_i(q)$ is also given by Eq. (\ref{approx-dispersion})
with
\begin{eqnarray}
\nonumber
\Delta_1 &=& \Delta_4 = 2\nphi(2u_1 - u_0), \\
\label{approx-dispersion-pfm}
\Delta_2 &=& - \ez + \Delta_1, \;\;\;\; \Delta_3 = \ez + \Delta_1, \\
\nonumber
\rho^i_S &=& \nphi u_1/4\pi + \rho^0_S,
\;\;\;\;\;\;\;\;\;\;\;\;\;\;\;\;\;\;\;\; i=1,2,3,4.
\end{eqnarray}
Here the branches $i=1$ and $4$ describe pseudospin wave excitations,
while $i=2$ and $3$ correspond to mixed spin-pseudospin wave
excitations [see Fig. \ref{fig03}].
The boson-boson interaction potential in $\bar{\mathcal{H}}_I^B$ is
\begin{eqnarray}
\nonumber
 \bar{v}^{i,j}_\bq(\bk,\bp) &=& e^{-(lq)^2/2}
    \left(u_0\sin\left(\bq\wedge\bk/2\right)\sin\left(\bq\wedge\bp/2\right) \right.
\\ \nonumber \\ \nonumber
          &+& \left. (u_0 - 4u_1)\cos\left(\bq\wedge\bk/2\right)\cos\left(\bq\wedge\bp/2\right)
          \right) \\
\nonumber \\ \nonumber
           &-& \delta_{i,j}\frac{u_0}{3}e^{-(lq)^2/2}\cos\left(\bq\wedge(\bk - \bp)/2\right)
\\ \nonumber \\ \nonumber
               &+& \bar{\delta}_{i,j}\frac{u_0}{3}\left[
               h_{i,j}e^{-(lq)^2/2}\cos\left(\bq\wedge(\bk - \bp)/2\right)
               \right.\\
\nonumber \\ \nonumber
          &+& \left. \bar{h}_{i,j}\left(i(-1)^{i+j}e^{-(lq)^2/2}
                     \sin\left(\bq\wedge(\bk - \bp)/2\right) \right.\right. \\
\nonumber \\
              &-&  \left.\left. e^{-l^2|\bq + \bk -\bp|^2/2}e^{-i(-1)^{i+j}\bq\wedge(\bk -
              \bp)/2}\right)
               \right]
\end{eqnarray}
Finally, the interaction term $\mathcal{V}^B$ can be written as
\begin{eqnarray}
\nonumber
\mathcal{V}^B &=&
       \sum_{\bq,\bp,\bk}v"_\bq(\bk,\bp)\left(
       b^\dagger_1(\bk+\bq)b^\dagger_4(\bp-\bq)b_3(\bp)b_2(\bk)\right.\\
                &+&\left. b^\dagger_2(\bk+\bq)b^\dagger_3(\bp-\bq)b_4(\bp)b_1(\bk)
                \right),
\label{int-boson-model3}
\end{eqnarray}
with
\begin{eqnarray}
v"_\bq(\bk,\bp) &=& -\frac{2u_0}{3}\left( e^{-(lq)^2/2} + e^{-(l|\bq+\bk-\bp|)^2/2} \right)
\nonumber \\
                && \times e^{-i\bq\wedge(\bp-\bk)/2}.
\label{coef-int-boson-model3}
\end{eqnarray}

The small and large momentum expansions of Eqs.
(\ref{dispersion-nu01}) and (\ref{dispersion-nu02}) are in agreement
with the results of Alicea and Fisher,\cite{alicea06} who present a
detailed discussion about the stability of each phase. Here we just
want to point out that the behavior of the smallest excitation gap
indicates which phase should set in. For instance, in the spin phase,
$\bar{\omega}_{2}(q=0)$ and $\bar{\omega}_{3}(q=0)$ are smaller than
$\bar{\omega}_{1}(q=0)$ and $\bar{\omega}_{4}(q=0)$ as long as $u_0 -
2u_1 < 0$. This result implies that the spin phase is stable only if
\[
  0 < \bar{\omega}_{2,3}(q=0) = \ez + 2\nphi(u_0 - 2u_1),
\]
where the estimated values of the parameters $\ez$, $u_0$, and $u_1$
are shown in the Appendix \ref{appendix1}. It is possible to show that
the spin phase sets in only if $U > U_C \sim 3.25\, eV$. The opposite
condition is found by carrying out the same analysis in the pseudospin
phase. It is difficult to conclude which phase is more favorable due
to the uncertainties in the determination of the on-site repulsion
term $U$.

{\it Charged excitations --} Concerning the elementary charged
excitations, the similarities between the effective boson model
derived from $\mathcal{H}_{SU4}$ [Eq. (\ref{hsu4})] and the SU(2)
counterpart\cite{doretto} indicate that, in both phases, the lowest
energy charged excitations should be described by quantum Hall
skyrmions as well. Again, within our formalism, such kind of
topological excitation is given by the state (\ref{skyrmion}). This
scenario agrees with the numerical calculations of Yang {\it et
al.},\cite{yang06} who show that skyrmions should occur in the $n=0$
as well as $n=1,\;2,\; {\rm and}\; 3$ Dirac Landau levels.

We can carry out the same analysis of the previous section and
calculate the semiclassical limit of the corresponding SU(2) boson
models for the spin and pseudospin phases in order to estimate how
$\mathcal{H}_{SB}$ changes the skyrmion energy. It is easy to show
that the energy functional $E_i[\mathbf{n}(\br)]$ is as in Eq.
(\ref{sondhi-models1}). For the spin phase, the parameters of
$E_i[\mathbf{n}(\br)]$ are
$\tilde{u}_1 = \tilde{u}_4 = 0$,
$\tilde{u}_2 = \tilde{u}_3 = 32\nphi(u_0/6 - u_1)$,
$\tilde{v}_1 = \tilde{v}_4 = 2\nphi(2u_0/3 - u_1)$,
$\tilde{v}_2 = \tilde{v}_3 = \nphi u_0/3$,
and  $\Delta_i$ and $\rho^i_S$ are given by Eqs.
(\ref{approx-dispersion-sfm}). Again, if $U \approx 10\,eV$, the
corrections due to the SU(4) symmetry breaking terms are such that all
stiffnesses are equal and so the topological-charge-topological-charge
interaction potential. In this case, we also have $\Delta_1 = \Delta_4
\approx \Delta_2 = \Delta_3$, and therefore both mixed spin-pseudospin
and spin skyrmions can be realized.

For the pseudospin phase, $E_i[\mathbf{n}(\br)]$ is characterized by
$\tilde{u}_1 = \tilde{u}_2 = \tilde{u}_3 = \tilde{u}_4 = 32\nphi(u_0/6
- u_1)$, $\tilde{v}_1 = \tilde{v}_2 = \tilde{v}_3 = \tilde{v}_4 =
\nphi u_0/3$, and $\Delta_i$ and $\rho^i_S$ given by Eqs.
(\ref{approx-dispersion-pfm}). Here the SU(4) symmetry breaking terms
equally affect the four excitation branches, independently of the
value of the on-site energy $U$. The mixed spin-pseudospin skyrmions
should have a larger radius (and therefore lower energy) than the
pseudospin ones because $\Delta_2/\epsilon_c < \Delta_1/\epsilon_c$
and $\Delta_4/\epsilon_c$ as it was already pointed out in Ref.
\onlinecite{alicea06}. The term $E^{ZZ}[\mathbf{n}(\br)]$ will even
reduce the energy of the texture with larger radius.

Finally, we should emphasize that our bosonization method is designed
to study {\it only} bulk excitations. The recent proposal of Abanin
and co-workers\cite {abanin07} that the finite value of the
longitudinal conductivity at $\nu=0$ is related to the existence of
charged gapless excitations at the {\it edge} of the system can not be
addressed with our formalism.

\section{Summary}

We presented here a non-perturbative bosonization scheme for electrons
restricted to the lowest Landau level in the presence of two discrete
degrees of freedom, spin-1/2 and pseudospin-1/2. We analyzed the cases
when the lowest Landau level is quarter-filled and half-filled. For
the latter, two distinct phases can be realized, the so-called spin
and pseudospin phases whereas in the former only the spin-pseudospin
phases sets in. In each case, a set of $n$-independent kinds of
creation and annihilation boson operators were defined and the boson
representation of the projected electron, spin, pseudospin, and mixed
spin-pseudospin density operators were calculated. The bosonic
expressions derived obey the lowest Landau level algebra.

We then applied the formalism to study the QHE at $\nu=0$ and $\nu=-1$
in graphene. We concentrated on very clean samples, assuming that the
system is in the quantum Hall ferromagnetic regime. For each quantum
Hall state, the continuous fermionic model proposed by Alicea and
Fisher\cite{alicea06} was mapped into an effective interacting boson
model. We showed that the quadratic term of this model describes $n$
well-defined branches of bosonic excitations, whose dispersion
relations are in agreement with the asymptotic ones calculated by
Alicea and Fisher.\cite{alicea06} Our formalism allows us to go beyond
the analysis presented in Ref.\ \onlinecite{alicea06} as we are able
to calculate the interaction between the $n$ bosonic excitation
branches.

The boson model $\mathcal{H}^B_{SU4}$ derived from the SU(4) invariant
part of the fermionic Hamiltonian is similar to its SU(2) counterpart
obtained before in our studies of the 2DEG at $\nu=1$. Based on this
analogy, we argued that the charged excitations for the quantum Hall
states in graphene should be describe by topological solitons (quantum
Hall skyrmions) and proposed that such excitation can be written as a
bosonic coherent state $|\rm sk\rangle$, generalizing the SU(2)
expression of Ref. \onlinecite{doretto}. We then calculated the
semiclassical limit of $\mathcal{H}^B_{SU4}$ and showed that the
derived energy functional is equal to the one calculated by Arovas
{\it et al.} for SU(N) quantum Hall skyrmions.\cite{arovas99}

We briefly discussed how the SU(4) symmetry breaking terms modify the
skyrmion energy functional by taking SU(2) limits of the total boson
model and then calculating the semiclassical limit of the reduced
models, i.e., focusing on one specific skyrmion flavor. We showed that
both the stiffness and the topological-charge-topological-charge
interaction potential are renormalized and that an extra term
($E_i^{ZZ}[\mathbf{n}]$), which favors larger skyrmions radius, is
introduced. More detailed studies of the boson-boson interaction
potential as well as of the disorder effects are deferred to a later
publication.

The method presented here is quite general. It might be used to study
bilayer quantum Hall systems at $\nu_T=1$ and $\nu_T=2$ realized in
GaAs heterostructures. In particular, it will allows us to address
questions related to the electronic spin, which seems to play an
important role in the behavior of these systems at $\nu_T=1$.

\begin{acknowledgments}
We are very grateful for the discussions with A.\ O. Caldeira, M.\ O.\
Goerbig and Ant\^onio Castro Neto. R.L.D. kindly acknowledges the
financial support from Conselho
Nacional de Desenvolvimento Cient\'ifico e Tecnol\'ogico (CNPq) -
Brazil and Deutsche ForschungsGemeinschaf (DFG) through research
grant SFB608.
\end{acknowledgments}

\appendix

\section{Energy scales}
\label{appendix1}

The relevant energy scales for the QHE in graphene are presented
in Table \ref{tab}. The cyclotron ($\hbar w_C$), Zeeman ($\ez$),
and Coulomb energies ($\epsilon_C$) as well as the parameters
$\nphi u_0$ and $\nphi u_1$ [see Sec. \ref{alicea-model}] are
given in terms of the magnetic field B, which is measured in
Tesla. We consider the following estimated parameters for
graphene: effective $g$-factor $g \approx 2$, dielectric constant
$\epsilon \approx 5$, and on-site repulsion energy $2 < U <
12\;eV$.\cite{alicea06} The magnetic length $l = \sqrt{\hbar
c/(eB)} = 256/\sqrt{B}$ is measured in angstroms and $a_0 =
2.46\,\AA$ is the lattice spacing of the triangular underlining
lattice.

\begin{table}[h]
\label{tab}
\caption{Energy scales for the QHE in graphene}
\begin{tabular}{l|rrrr}\hline
Energy scales & \hskip1.5cm &    &  \hskip1.0cm            & (K)    \\ \hline
$\hbar w_C$   &               &$\sqrt{2\hbar v^2_F B/c}$ &     & $380.60\sqrt{B}$ \\
$ \ez $       &               & $g\mu _BB$     &               & $ 1.08B$ \\
$\epsilon_C$  &               &$e^2/\epsilon l$&               & $150.12\sqrt{B}$\\
$\nphi u_0$   &               &$\sqrt{3}Ua^2_0/8\pi l^2$&      & $0.08UB$\\
$\nphi u_1$   &               &$a_0l\epsilon_C/\sqrt{3}$&      & $0.4B$\\
\hline
\end{tabular}
\end{table}

\section{Boson representation of the density operators $\rho_{12}(\bq)$
  $\rho_{21}(\bq)$, $\rho_{34}(\bq)$, and $\rho_{43}(\bq)$}
\label{appendix2}

Let us concentrate on the spin-pseudospin phase. Although the
boson operators $b_1(\bq)$ and $b^\dagger_1(\bq)$ are respectively
defined by $\rho_{12}(\bq)$ and $\rho_{21}(\bq)$, the boson
representations of these density operators are not necessarily
$\rho_{12}(\bq) = \alpha_\bq^{-1}b_1(-\bq)$ and $\rho_{21}(\bq) =
\alpha_\bq^{-1}b^\dagger_1(\bq)$. If it were the case, we would
have $[\rho_{12}(\bq),\rho_{21}(\bk)] =
\alpha_\bq^{-2}\delta_{\bq,-\bk}$, in completely disagreement with
the commutator (\ref{commutator}). The same procedure described in
Sec. \ref{spfm-phase} should be employed in this case as well. Due
to the similarities between the steps involved here and in the
calculation of the boson representation of the spin density
operators $S^+(\bq)$ and $S^-(\bq)$ of the SU(2) case, we refer
the reader to Sec. II.C of Ref.\ \onlinecite{doretto} for all the
details and just display the final results here. We have,
\begin{eqnarray}
\nonumber
\rho_{21}(\bq) &\equiv& \sqrt{\nphi}e^{-(lq)^2/4}b^\dagger_1(\bq), \\
\label{ops1-spfm} && \\
\nonumber
\rho_{12}(\bq) &=& \sqrt{\nphi}e^{-(lq)^2/4}b_1(-\bq) \\
\nonumber && \\
\nonumber
          &-&\sum_{i,\bk,\bp}f_i^{12}(\bq,\bk,\bp)b^\dagger_i(\bk+\bq+\bp)b_i(\bp)b_1(\bk),
\end{eqnarray}
where the form factors are given by
\begin{eqnarray}
\nonumber f_1^{12}(\bq,\bk,\bp) &=&
 \nphi^{-1/2}e^{-(lq)^2/4}\cos((\bq+\bk)\wedge(\bp+\bq)/2), \\
\nonumber && \\
\nonumber
f_2^{12}(\bq,\bk,\bp) &=& f_3^{12}(\bq,\bk,\bp)\\
\nonumber && \\
\nonumber
  &=& \nphi^{-1/2}e^{-(lq)^2/4}e^{-i(\bq+\bk)\wedge(\bp+\bq)/2},
\end{eqnarray}
and
\begin{eqnarray}
\nonumber
\rho_{34}(\bq)&=& \rho^*_{34}(-\bq) \\
\nonumber && \\
              &=& e^{-(lq)^2/4}\sum_\bk e^{i\bq\wedge\bk/2}
                  b^\dagger_2(\bq+\bk)b_3(\bk).
\label{ops2-spfm}
\end{eqnarray}

Similar considerations hold for the spin phase. In this case, the
boson representation of both $\rho_{21}(\bq)$ and $\rho_{43}(\bq)$
are defined respectively by the creation boson operators
$b^\dagger_1(\bq)$ and $b^\dagger_4(\bq)$, while more involve
expressions are derived for their Hermitian conjugates. We have,
\begin{eqnarray}
\nonumber
\rho_{21}(\bq) &\equiv& \sqrt{\nphi}e^{-(lq)^2/4}b^\dagger_1(\bq), \\
\label{ops1-sfm} && \\
\nonumber
\rho_{12}(\bq) &=& \sqrt{\nphi}e^{-(lq)^2/4}b_1(-\bq) \\
\nonumber && \\
\nonumber
          &-&\sum_{i=1}^3\sum_{\bk,\bp}f_i^{12}(\bq,\bk,\bp)
          b^\dagger_i(\bk+\bq+\bp)b_i(\bp)b_1(\bk)\\
\nonumber && \\
\nonumber
          &-&\sum_{\bk,\bp}\bar{f}^{12}(\bq,\bk,\bp)b^\dagger_4(\bk+\bq+\bp)b_3(\bp)b_2(\bk),
\end{eqnarray}
where the form factors are
\begin{eqnarray}
\nonumber
f_1^{12}(\bq,\bk,\bp) &=&
        \nphi^{-1/2}e^{-(lq)^2/4}\cos((\bq+\bk)\wedge(\bp+\bq)/2), \\
\nonumber && \\
\nonumber
f_2^{12}(\bq,\bk,\bp) &=& \bar{f}^{12}(\bq,\bk,\bp) \\
\nonumber && \\
\nonumber
        &=& \nphi^{-1/2}e^{-(lq)^2/4}e^{+i(\bq+\bk)\wedge(\bp+\bq)/2}, \\
\nonumber && \\
\nonumber
f_3^{12}(\bq,\bk,\bp) &=&
        \nphi^{-1/2}e^{-(lq)^2/4}e^{-i(\bq+\bk)\wedge(\bp+\bq)/2},
\end{eqnarray}
and
\begin{eqnarray}
\nonumber
\rho_{43}(\bq) &\equiv& \sqrt{\nphi}e^{-(lq)^2/4}b^\dagger_4(\bq), \\
\label{ops2-sfm} && \\
\nonumber
\rho_{34}(\bq) &=& \sqrt{\nphi}e^{-(lq)^2/4}b_4(-\bq) \\
\nonumber && \\
\nonumber
          &-&\sum_{i=2}^4\sum_{\bk,\bp}f_i^{34}(\bq,\bk,\bp)
          b^\dagger_i(\bk+\bq+\bp)b_i(\bp)b_4(\bk)\\
\nonumber && \\
\nonumber
          &-&\sum_{\bk,\bp}\bar{f}^{34}(\bq,\bk,\bp)b^\dagger_1(\bk+\bq+\bp)b_2(\bp)b_3(\bk),
\end{eqnarray}
with the following form factors
\begin{eqnarray}
\nonumber
f_2^{34}(\bq,\bk,\bp) &=& \bar{f}^{34}(\bq,\bk,\bp) \\
\nonumber && \\
\nonumber
          &=& \nphi^{-1/2}e^{-(lq)^2/4}e^{-i(\bq+\bk)\wedge(\bp+\bq)/2}, \\
\nonumber && \\
\nonumber
f_3^{34}(\bq,\bk,\bp) &=&
        \nphi^{-1/2}e^{-(lq)^2/4}e^{+i(\bq+\bk)\wedge(\bp+\bq)/2}, \\
\nonumber && \\
\nonumber
f_4^{34}(\bq,\bk,\bp) &=&
        \nphi^{-1/2}e^{-(lq)^2/4}\cos((\bq+\bk)\wedge(\bp+\bq)/2).
\end{eqnarray}

Finally, the pseudospin phase is characterized by the following
expressions
\begin{eqnarray}
\nonumber
\rho_{12}(\bq) &=& - e^{-(lq)^2/4}\sum_\bk e^{-i\bq\wedge\bk/2}
                   \left(b^\dagger_2(\bq+\bk)b_1(\bk)\right. \\
\nonumber && \\
\nonumber      && + \left. b^\dagger_4(\bq+\bk)b_3(\bk)\right) \\
\label{ops-pfm} && \\
\nonumber
\rho_{34}(\bq) &=& e^{-(lq)^2/4}\sum_\bk e^{i\bq\wedge\bk/2}
                   \left(b^\dagger_1(\bq+\bk)b_3(\bk) \right. \\
\nonumber && \\
\nonumber      && + \left. b^\dagger_2(\bq+\bk)b_4(\bk)\right),
\end{eqnarray}
with $\rho_{21}(\bq) = \rho^*_{12}(-\bq)$ and $\rho_{43}(\bq) =
\rho^*_{34}(-\bq)$. The derivation of the above expressions is
analogous to the one involved in the calculations of Eq. (\ref{ops2-spfm}).

\section{Bosonization and Dirac Landau levels}
\label{appendix3}

For electrons in graphene subject to a perpendicular magnetic
field, the fermion field operator is a two-component spinor, which
may be written in Dirac Landau level basis as
\begin{eqnarray}
\hat{\Psi}^\dagger_{\alpha\,\sigma}(\br) &=&
       e^{-i\alpha\mathbf{K}\cdot\br}\sum_{n,m} \langle
       \hat{\Phi}_{n\, m\,\alpha}|\br\rangle
       c^\dagger_{n\, m\, \alpha\, \sigma}, \nonumber \\
&& \label{spinor-fermionfields} \\
\hat{\Psi}_{\alpha\,\sigma}(\br) &=&
      e^{i\alpha\mathbf{K}\cdot\br}\sum_{n,m} \langle
      \br | \hat{\Phi}_{n\, m\,\alpha} \rangle
      c_{n\, m\, \alpha\, \sigma},
\nonumber
\end{eqnarray}
where $\mathbf{K} = (2\pi/a_0)(2/3,0)$ and
the spinors $| \hat{\Phi}_{n\, m\,\alpha} \rangle$ are given by
Eqs. (\ref{ll-eigenstates1}) and (\ref{ll-eigenstates2}).

Defining the density operator
$\hat{\rho}_{\alpha\sigma,\beta\lambda}(\br)$ as
\begin{equation}
\hat{\rho}_{\alpha\sigma,\beta\lambda}(\br) =
\hat{\Psi}^\dagger_{\alpha\,\sigma}(\br)\hat{\Psi}_{\beta\,\lambda}(\br),
\end{equation}
one can calculate its Fourier transform in the same way as it is
done in Eq. (\ref{fourier}), i.e.,
\begin{eqnarray}
\nonumber
\hat{\rho}_{\alpha\sigma,\beta\lambda}(\bq) &=&
         \sum_{n,n'}\sum_{m,m'}
         \langle \hat{\Phi}_{n\, m\,\alpha}|
          e^{-i\left(\bq + (\alpha - \beta)\mathbf{K}\right)\cdot\br}|
          \hat{\Phi}_{n'\, m'\,\beta} \rangle  \\
\nonumber && \\
          && \times c^\dagger_{n\, m\, \alpha\, \sigma}c_{n'\, m'\, \beta\,
          \lambda}.
\label{fourier2}
\end{eqnarray}
The projection into the $n$-th Dirac Landau level is obtained by
taking the component $n=n'$ in Eq. (\ref{fourier2}). For the lowest
Landau level, the fact that the eigenvectors (\ref{ll-eigenstates2})
have only one non-zero entry implies that
\begin{eqnarray}
\nonumber
\bar{\rho}_{\alpha\sigma,\beta\lambda}(\bq) &=& e^{-(lq)^2/2}F^{\alpha\beta}_{n=0} \\
\nonumber && \\
\nonumber
     && \times\sum_{m,m'}G_{m,m'}(l\bq)c^\dagger_{0\, m\, \alpha\,
       \sigma}c_{0\, m'\, \beta\, \lambda}, \\
\nonumber && \\
     &=&F^{\alpha\beta}_{n=0}\rho_{\alpha\sigma,\beta\lambda}(\bq),
\label{fourier4}
\end{eqnarray}
where 
$F^{\alpha\beta}_{n=0}=\delta_{\alpha,\beta}$ and
$\rho_{\alpha\sigma,\beta\lambda}(\bq)$ is given by Eq.
(\ref{fourier}). In the isospin language, $\bar{\rho}_{IJ}(\bq) =
\rho_{IJ}(\bq)$ and does not vanish only if $(I,J) =
(I,I),\,(1,2),\,(2,1),\,(3,4),\;{\rm and}\;(4,3)$. Notice that
these are the density operators which appear in the effective
continuous model (\ref{continuous-model}). Therefore, the
expressions derived in Secs. \ref{spfm-phase} - \ref{pfm-phase}
can be directly employed to study the fermionic model
(\ref{continuous-model}).

The situation is more complex for higher Landau levels
($n\not=0$). From Eqs. (\ref{ll-eigenstates1}), it is possible to show
that\cite{goerbig06}
\begin{eqnarray}
\nonumber
\bar{\rho}_{\alpha\sigma,\beta\lambda}(\bq) &=&
   e^{-|l\bq - l\mathbf{K}|^2/2}F^{\alpha\beta}_{n}(\bq) \\
\nonumber && \\
\nonumber
     && \times\sum_{m,m'}G_{m,m'}(l\bq - l\mathbf{K})c^\dagger_{n\, m\, \alpha\,
       \sigma}c_{n\, m'\, \beta\, \lambda}, \\
\nonumber && \\
     &=&F^{\alpha\beta}_{n}(\bq)\rho_{\alpha\sigma,\beta\lambda}(l\bq
     + l\mathbf{K}), \label{fourier3}
\end{eqnarray}
where the form factors $F^{\alpha\beta}_{n}(\bq)$ read
\begin{eqnarray}
\nonumber
F^{\alpha\alpha}_{n}(\bq) &=& \frac{1}{2}\left[L_{|n|}\left(\frac{(lq)^2}{2}\right)
                              + L_{|n|-1}\left(\frac{(lq)^2}{2}\right)\right], \\
&& \\
\nonumber F^{+-}_{n}(\bq) &=& \frac{1}{\sqrt{2n}}\left(lq_x -
                     lK_x\right)
                     L^1_{|n|-1}\left(\frac{|l\bq -
                     l\mathbf{K}|^2}{2}\right),
\end{eqnarray}
and $F^{-+}_{n}(\bq) = F^{+-}_{n}(-\bq)$ with $\bkp \rightarrow -\bkp$. In the expressions
above, we used the fact that $2\mathbf{K} = \mathbf{K'} =
-\mathbf{K}$. The connection with the formulae derived in Secs.
\ref{spfm-phase} - \ref{pfm-phase} is obtained via the relation.
\begin{equation}
\bar{\rho}_{\alpha\sigma,\beta\lambda}(l\bq - l\mathbf{K})
   =F^{\alpha\beta}_{n}(l\bq -
   l\mathbf{K})\rho_{\alpha\sigma,\beta\lambda}(l\bq).
\end{equation}
We refer the reader to Ref. \onlinecite{goerbig06} for a detailed
analysis of the form factors $F^{\alpha\beta}_{n}(\bq)$ and theirs
implications in the dynamics of the quantum Hall states in higher
Dirac Landau levels.


\begin{thebibliography}{99}
\bibitem{nota} In order to be consistent with the previous sections,
               we should, in principle, add  $-\ep P_Z(\bq=0)$ to the total
               Hamiltonian because it does not have any one-particle term
               which explicitly breaks the pseudospin symmetry. After
               substituting each density operator by its bosonic
               representation and normal ordering the result, we can
               take the limit $\ep\rightarrow 0$.
\end{thebibliography}
\end{document}